\documentclass[11pt,a4paper]{article}
\pdfoutput=1
\usepackage{jcappub}	

\usepackage[normalem]{ulem}	

\newcommand{\blue}[1]{\color{blue} #1 \color{black}}

\usepackage{graphicx}
\usepackage{multirow}

\usepackage{lineno}
\usepackage{cancel}
\usepackage{here}

\def\lsim{\raise0.3ex\hbox{$\;<$\kern-0.75em\raise-1.1ex
\hbox{$\sim\;$}}}
\def\gsim{\raise0.3ex\hbox{$\;>$\kern-0.75em\raise-1.1ex
\hbox{$\sim\;$}}}

\title
{
\blue{How Unequal Fluxes of High Energy Astrophysical Neutrinos 
and Antineutrinos can Fake New Physics}
}
\author{Hiroshi Nunokawa$^{1}$}
\author{Boris Panes$^{2}$}
\author{and Renata Zukanovich Funchal$^{2}$}

\affiliation{
$^1$Departamento de F\'{\i}sica, Pontif{\'\i}cia Universidade Cat{\'o}lica 
do Rio de Janeiro, C. P. 38071, 22452-970, Rio de Janeiro, Brazil  \\
$^2$Instituto de F\'{\i}sica, Universidade de S\~ao
  Paulo, C.\ P.\ 66.318, 05315-970 S\~ao Paulo, Brazil 
}
\emailAdd{nunokawa@puc-rio.br}
\emailAdd{bapanes@if.usp.br} 
\emailAdd{zukanov@if.usp.br}

\abstract{Flavor ratios of very high energy astrophysical neutrinos,
  which can be studied at the Earth by a neutrino telescope such as
  IceCube, can serve to diagnose their production mechanism at the
  astrophysical source.  The flavor ratios for neutrinos and
  antineutrinos can be quite different as we do not know how they are
  produced in the astrophysical environment.  Due to this uncertainty
  the neutrino and antineutrino flavor ratios at the Earth also could
  be quite different.  Nonetheless, it is generally assumed that
  flavor ratios for neutrinos and antineutrinos are the same at the
  Earth, in fitting the high energy astrophysical neutrino data.  This
  is a reasonable assumption for the limited statistics for the data
  we currently have.  However, in the future the fit must be performed
  allowing for a possible discrepancy in these two fractions in order
  to be able to disentangle different production mechanisms at the
  source from new physics in the neutrino sector. To reinforce this
  issue, in this work we show that a wrong assumption about the
  distribution of neutrino flavor ratios at the Earth may indeed lead
  to misleading interpretations of IceCube results.  }


\begin{document}
\maketitle

\section{Introduction}
\label{sec:intro}

The discovery by the IceCube Collaboration that the Earth is bombarded
by a flux of very high energy extraterrestrial
neutrinos~\cite{Aartsen:2013bka,Aartsen:2013jdh,Aartsen:2014gkd}, 
opened a new research field, High Energy Neutrino Astronomy. 
Naturally, after these findings a big amount of activity has been
devoted to studies about the origin of these neutrinos, see for instance
\cite{Bhattacharya:2011qu,Esmaili:2013gha,Fong:2014bsa,Roulet:2012rv,Cholis:2012kq,Kalashev:2013vba,Stecker:2013fxa,Murase:2013rfa}. 
Furthermore, as more data is collected it also becomes possible to study finer details about the observed neutrinos, which has motivated many studies devoted to address the flavor composition of this
flux~\cite{Mena:2014sja,Bustamante:2015waa,Palladino:2015zua,Palomares-Ruiz:2015mka,Watanabe:2014qua,Kawanaka:2015qza,Aartsen:2015knd,Chen:2014gxa}.
The reason to focus on composition studies relies on the fact that
neutrino flavor composition can be used to tackle the production
mechanism of very high energy neutrinos at the
source, as well as to shed light on the possible presence 
of non-standard neutrino properties beyond 
the standard three flavor mixing scheme.

Understanding this process is utterly momentous as the same mechanism is expected to
be at the genesis of very-high energy cosmic rays and
photons~\cite{Waxman:1997ti,Waxman:1998yy,Bento:1999bb,Bechtol:2015uqb,Gaggero:2015nga,Kadler:2016ygj,Moharana:2016xkz,Aartsen:2016qcr,Padovani:2016wwn,Aartsen:2015dml,Gaggero:2015xza,Lunardini:2011br,Lunardini:2015laa,Aloisio:2015ega,Murase:2015gea,Murase:2015xka,Hummer:2010ai,Hummer:2010vx}. 
We hope correlations among these ultra-high energy cosmic particles can
be established in the near future helping us to unravel this long standing  mystery.  
In the most frequently considered scenario, the
decay of pions and subsequently of muons dominate the neutrino flux,
producing the flux flavor ratios 
($f^{\rm S}_{\nu_e}$ :
$f^{\rm S}_{\nu_\mu}$:
$f^{\rm S}_{\nu_\tau}$)\footnote{Notice that here we are using italic $f$ for the fractions. This is the notation that we use when we consider equal fractions of neutrinos and antineutrinos. Later, we will introduce a notation where fractions are specified with a normal f, in order to consider neutrinos and 
antineutrinos independently.}=$(\frac{1}{3} : \frac{2}{3} : 0)_{\rm S}$ at the
source~\cite{Learned:1994wg}.
However, the flux flavor composition at
the source may vary from $(1 : 0 : 0)_{\rm S}$ to $(0 : 1 : 0)_{\rm S}$ under a
profusion of different scenarios which include muon energy
loss~\cite{Rachen:1998fd,Kashti:2005qa,Kachelrie[]:2006fi,Lipari:2007su},
muon acceleration~\cite{Klein:2012ug}, and neutron
decay~\cite{Anchordoqui:2003vc}.

\begin{figure}[h!]
\centering
\begin{minipage}{1.0\textwidth}
\begin{minipage}{.5\textwidth}
\centering
\includegraphics[width=1.1\linewidth]{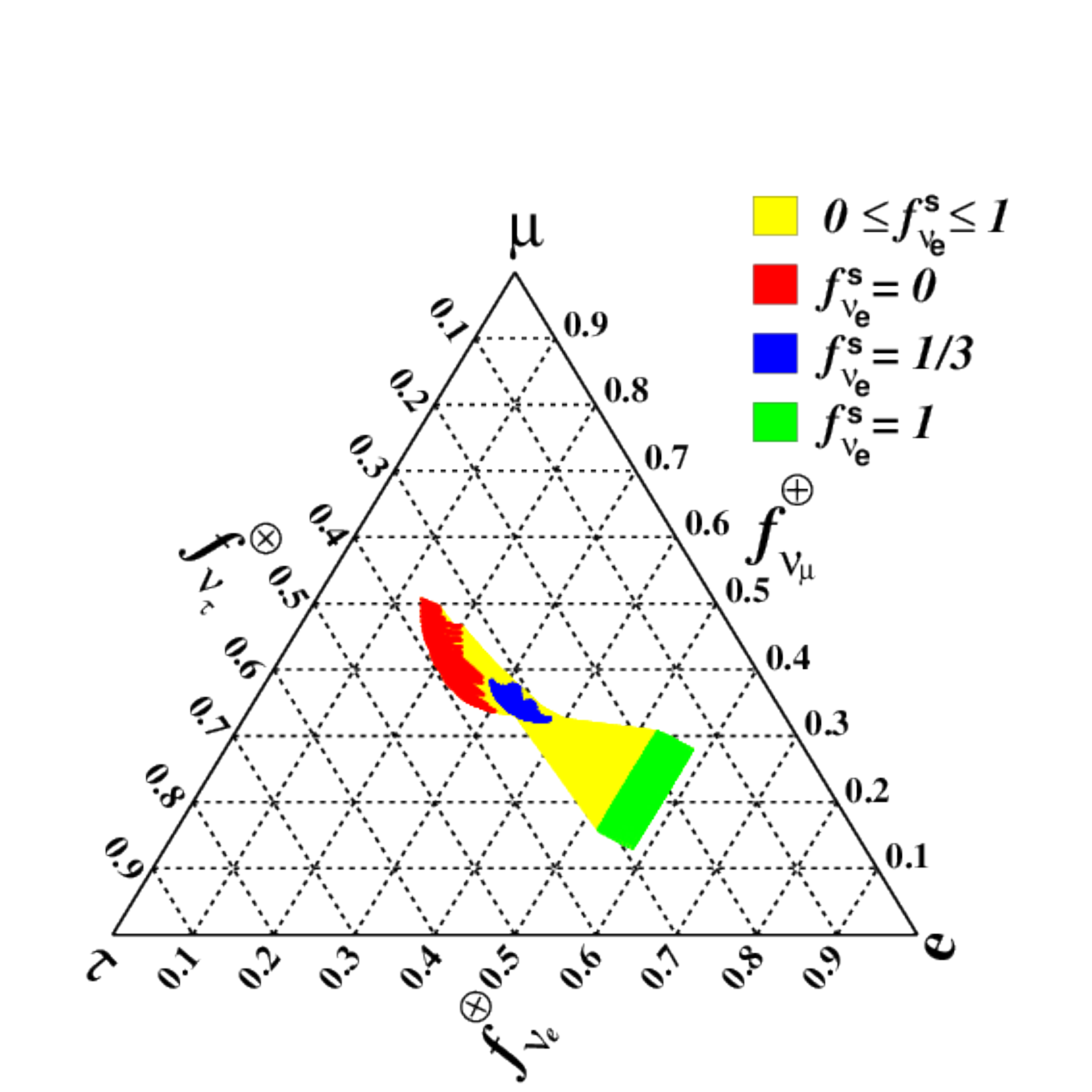}
\end{minipage}
\begin{minipage}{.5\textwidth}
\centering
\includegraphics[width=1.1\linewidth]{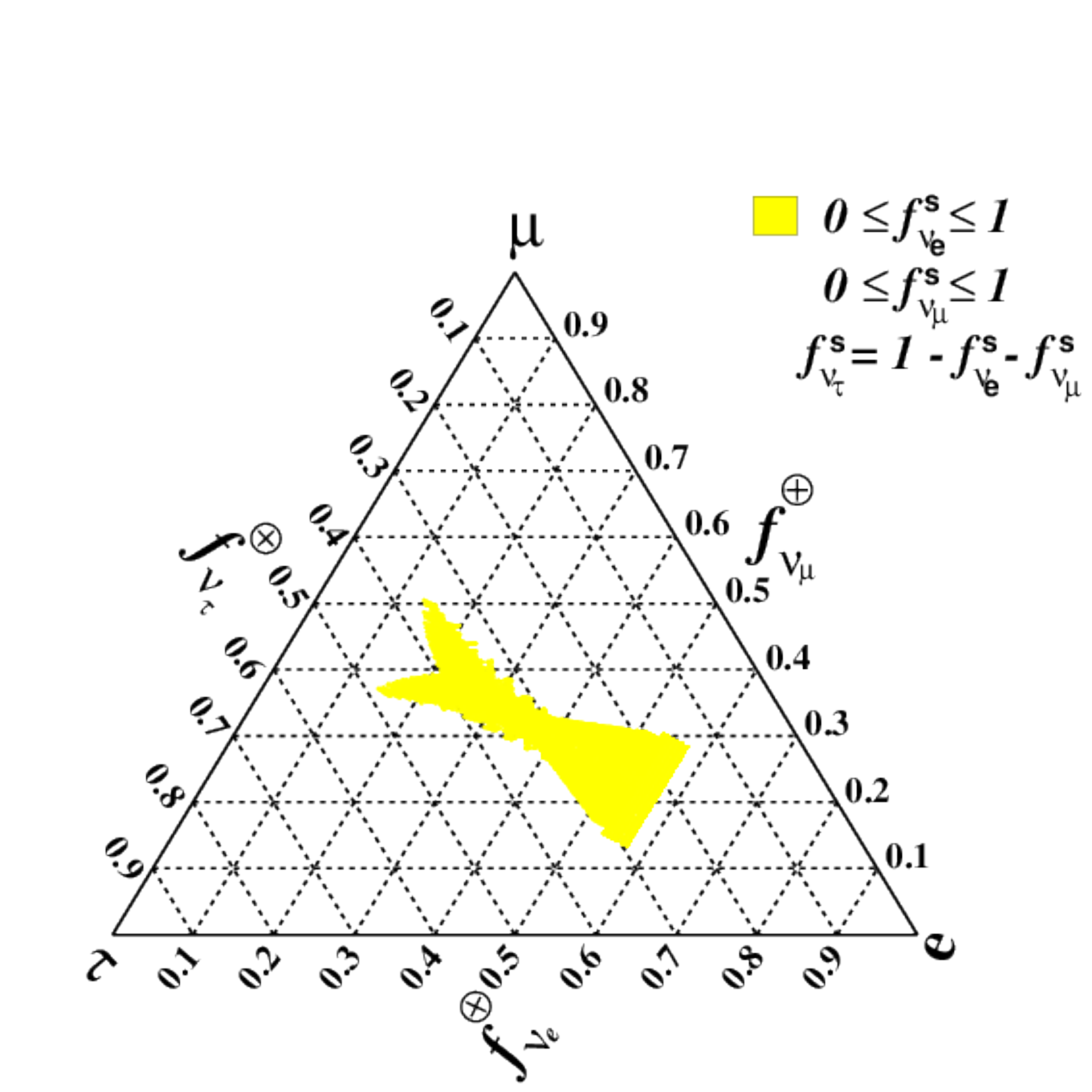}
\end{minipage}
\end{minipage}
\caption{\label{fig1:triangle-standard-expected}
Currently allowed (expected) regions for the flux flavor ratios 
($f^\oplus_{\nu_e}$, $f^\oplus_{\nu_\mu}$, $f^\oplus_{\nu_\tau}$) at the Earth 
by various high energy neutrino production scenarios denoted by the values of the 
parameter $f^{\rm S}_{\nu_e}$ which lies somewhere between 0 and 1.
On the left panel $\nu_\tau$ are not allowed to be 
produced at the source, while on the right panel they are. 
We let neutrino mixing parameters vary within 
the current 3$\sigma$ allowed regions~\cite{Bergstrom:2015rba}.
Red, blue and green regions correspond to the case
where $f^{\rm S}_{\nu_e} = 0, 1/3$ and 1, respectively, 
whereas in the yellow regions, 
any value of $f^{\rm S}_{\nu_e} \in [0,1]$ is allowed.}
\end{figure}

After propagating over astronomical distances, as first pointed out by
the authors of Ref.~\cite{Learned:1994wg}, neutrino oscillations tend
to equalize the flux flavor ratios to 
($f^\oplus_{\nu_e}$: $f^\oplus_{\nu_\mu}$: $f^\oplus_{\nu_\tau}$) 
$\approx (\frac{1}{3}:\frac{1}{3}:\frac{1}{3})_\oplus$ at the Earth, 
for the expected source composition $(\frac{1}{3}:\frac{2}{3}:0)_{\rm S}$.  
In fact, by varying the neutrino oscillation mixing parameters in the
regions allowed by the latest global fit~\cite{Bergstrom:2015rba} and
by assuming any possible source flavor composition, it can be shown
that the prediction for the flux flavor ratios at the Earth always end-up
somewhere around the central point
$(\frac{1}{3}:\frac{1}{3}:\frac{1}{3})_\oplus$, {\em i.e.}, in a
restricted region around the middle of 
the flavor-ratio triangle 
as can be seen on the left panel of Fig.~\ref{fig1:triangle-standard-expected} 
(see also \cite{Bustamante:2015waa,Arguelles:2015dca}).
This remains true
even if one allows for an arbitrary $\nu_\tau$ content at the source,
as depicted on the right panel of Fig.~\ref{fig1:triangle-standard-expected}, although one
predicts $\nu_\tau$ contribution to be quite
negligible~\cite{Athar:2005wg}.

\vglue -1.0cm

Because of the limited variation of the neutrino flavor ratios at the
Earth, some authors have speculated that a deviation of the neutrino
flux flavor ratios from this central region of 
the flavor-ratio triangle 
could be interpreted as an indication of new physics at play in the
neutrino sector. Among the other exotic possibilities, neutrino
decay~\cite{Beacom:2002vi,Baerwald:2012kc}, pseudo-Dirac
neutrinos~\cite{Beacom:2003eu,Esmaili:2009fk}, sterile
neutrinos~\cite{Athar:2000yw}, Lorentz or CPT
violation~\cite{Hooper:2005jp}, quantum
decoherence~\cite{Anchordoqui:2005gj,Hooper:2004xr}, the impact of
effective operators \cite{Arguelles:2015dca}, 
non-standard neutrino interactions with dark matter~\cite{deSalas:2016svi}, 
and so on, have been considered in the literature.

Moreover, although still statistically insignificant at the moment,
fits to the IceCube data seem to be favoring flavor ratios at the
Earth outside the allowed regions shown in Fig.~\ref{fig1:triangle-standard-expected}, 
close to the edges of the flavor-ratio
triangle~\cite{Aartsen:2015ivb,Aartsen:2015knd} that might signal a
conflict with the standard expectation.  If this is confirmed by
future data, can this be an unambiguous sign of new physics or can this
be explained by standard physics?

We show in this paper that in order to use neutrino flux flavor ratios
at the Earth to probe the various production mechanism scenarios at
the source, one cannot assume that neutrinos and antineutrinos fluxes
at the Earth are the same when fitting the data. In particular, this
assumption may lead to erroneous interpretation of the data in the
future, and may already be the reason why the present best fit point
for the flux flavor ratios seems to be outside the expected region.

In Sec.~\ref{sec:mechanisms} we briefly review the main high energy
neutrino/antineutrino production mechanisms at astrophysical sources
as well as the neutrino oscillation effect on their propagation to the
Earth.  In Sec.~\ref{sec:pgamma} we discuss how unequal neutrino and
antineutrino fluxes can affect the fitting of the shower and track
events measured by IceCube. In particular, the fit of IceCube data by
imposing the equality of neutrino and antineutrino flavor ratios at
the Earth, when they are predicted to be not equal at the source and
therefore at the Earth also, can produce allowed regions in the
  flavor-ratio triangle that lay outside
  the one expected by standard oscillation and consequently be
  misinterpreted as a sign of new physics.  Finally, in
  Sec.~\ref{sec:conclusions} we provide some
    discussions and our conclusions.

\section{Neutrino and Antineutrino Production Mechanisms and Propagation}
\label{sec:mechanisms}

We do not know where or how exactly very high energy ($\gsim 100$ TeV)
astrophysical neutrinos are produced. It is generally believed that
their production takes place in transient astrophysical cosmic ray
engines such as blazars (jets in active galactic nuclei) and gamma-ray
bursts.  The general idea behind the main mechanisms is that protons
are accelerated and possibly confined in these sources by magnetic
fields. The production of neutrons and charged pions and their
succeeding decays produce neutrinos as well as other cosmic ray
particles. Neutrinos, in particular, are produced via $\pi^+ \to \mu^+
\nu_\mu \to e^+ \nu_e \overline{\nu}_\mu \nu_\mu$ (and its
charge conjugate process).

Depending on source properties such as the relative ambient gas and
photon densities, pion production proceeds either by inelastic 
$pp$ scattering~\cite{Anchordoqui:2003vc} or by photo disintegration,
primarily through the resonant processes $p \gamma \to \Delta^+ \to n
\pi^+, \, p \pi^0$~\cite{Waxman:1998yy}.

\subsection{\bf $pp$ production}
\label{subsec:pp-production}

If inelastic $pp$ collisions are the dominant process of pion
production, isospin invariance yields approximately equal numbers of
$\pi^{\pm}$.  Their decay followed by the decay of $\mu^\pm$ will lead
to the flux flavor ratios at the source of 

\begin{eqnarray}
(\text{f}^{\,\rm S}_{\nu_e}: \text{f}^{\,\rm S}_{\nu_\mu}:
 \text{f}^{\,\rm S}_{\nu_\tau})
\simeq 
(\text{f}^{\,\rm S}_{\overline{\nu}_e}:  
\text{f}^{\,\rm S}_{\overline{\nu}_\mu}: 
\text{f}^{\,\rm S}_{\overline{\nu}_\tau})
\simeq \left(\frac{1}{6}:\frac{2}{6}:0\right)_{\rm S}, 
\label{eq:flavor-ratio-normal-pp}
\end{eqnarray}
for both neutrinos and antineutrinos (note that sum of all fractions 
is normalized to 1).

Since in this case the flavor ratios are the same
for neutrinos and antineutrinos, one can simply talk about the
combined (neutrino+antineutrino) flavor ratios 

\begin{eqnarray}
({f}^{\,\rm S}_{\nu_e}: {f}^{\,\rm S}_{\nu_\mu}:
 {f}^{\,\rm S}_{\nu_\tau})
&\equiv &
(\text{f}^{\,\rm S}_{\nu_e}+\text{f}^{\,\rm S}_{\overline{\nu}_e}\, : \,  
\text{f}^{\,\rm S}_{\nu_\mu}+\text{f}^{\,\rm S}_{\overline{\nu}_\mu}\, : \, 
 \text{f}^{\,\rm S}_{\nu_\tau}+\text{f}^{\,\rm S}_{\overline{\nu}_\tau})
 \nonumber \\
&=&  
2(\text{f}^{\,\rm S}_{{\nu}_e} : \text{f}^{\,\rm S}_{{\nu}_\mu} : 
\text{f}^{\,\rm S}_{{\nu}_\tau})
= 
2(\text{f}^{\,\rm S}_{\overline{\nu}_e} :   
\text{f}^{\,\rm S}_{\overline{\nu}_\mu} :  
\text{f}^{\,\rm S}_{\overline{\nu}_\tau}) \nonumber \\
&\simeq &\left(\frac{1}{3}:\frac{2}{3}:0\right)_{\rm S}.
\label{eq:flavor-ratio-italic-pp}
\end{eqnarray}

Here, as we noted in the footnote 1, 
we have to distinguish two different notations: italic 
$f_{\nu_\alpha}$ for the case where we 
do not consider the neutrino and antineutrino 
flavor ratios independently, as they are {\em assumed} to be the same,
and normal f$_{\nu_\alpha}$  and f$_{\bar{\nu}_\alpha}$ for 
the case where we consider independently neutrino and antineutrino flavor ratios.

However, since the lifetime of the muon exceeds that of the charged
pion by almost two orders of magnitude, it is also possible that the
muon in the decay chain losses energy to the source environment by
synchrotron radiation and interaction before it
decays~\cite{Rachen:1998fd,Kashti:2005qa}.  
If this happens, these low-energy muons will have a negligible contribution to
the high energy neutrino flux. 
This will result in only $\nu_\mu$
and $\overline{\nu}_\mu$ produced at the source, {\em i.e.},

\begin{eqnarray}
(\text{f}^{\,\rm S}_{\nu_e}: \text{f}^{\,\rm S}_{\nu_\mu}:
\text{f}^{\,\rm S}_{\nu_\tau}) \simeq (\text{f}^{\,\rm S}_{\overline{\nu}_e}:
\text{f}^{\,\rm S}_{\overline{\nu}_\mu}:
\text{f}^{\,\rm S}_{\overline{\nu}_\tau}) \simeq 
\left(0:\frac{1}{2}:0\right)_{\rm S}. 
\end{eqnarray}

Again in this case we have the same flux for neutrinos and antineutrinos and we
can again simply use the combined flavor ratio ($f^{\,\rm S}_{\nu_e}$:
$f^{\,\rm S}_{\nu_\mu}$: 
$f^{\,\rm S}_{\nu_\tau}$) = (0 : 1 : 0)$_{\rm S}$.

\subsection{\bf $p\gamma$ production}
\label{subsec:pgamma-production}

On the other hand, if pions are produced mainly by 
$p\gamma$ interactions, the flux ratio of neutrinos and 
antineutrinos depends on the spectra of target photons,
as discussed in \cite{Shoemaker:2015qul}. 
If the target photons are very hard or thermal, 
multi-pion production would be more relevant and 
expected flavor ratio may become very similar to 
the $pp$ case, as in 
Eqs.~(\ref{eq:flavor-ratio-normal-pp}) and ~(\ref{eq:flavor-ratio-italic-pp}).
This is often expected in GRB and blazar models, 
see e.g. \cite{Murase:2013ffa,Murase:2014foa}.

However, if the target photons are sufficiently soft, 
the asymmetric process $p \gamma \to \Delta^+ \to n
\pi^+$ will be the dominant source of neutrinos.
Since $\pi^-$ production is suppressed 
and $\pi^+$ decays do not produce 
$\overline{\nu}_e$, this lead to the flux flavor ratios at the source as 
\begin{eqnarray}
(\text{f}^{\,\rm S}_{\nu_e}: \text{f}^{\,\rm S}_{\nu_\mu}: 
\text{f}^{\,\rm S}_{\nu_\tau}) 
\simeq 
\left(\frac{1}{3}:\frac{1}{3}:0\right)_{\rm S}, \  \ 
(\text{f}^{\,\rm S}_{\overline{\nu}_e}: \text{f}^{\,\rm S}_{\overline{\nu}_\mu}:
\text{f}^{\,\rm S}_{\overline{\nu}_\tau}) \simeq 
\left(0:\frac{1}{3}:0\right)_{\rm S}.
\label{eq:flavor-ratio-italic-pgamma}
\end{eqnarray}
In this case neutrinos and antineutrinos exhibit quite different
flavor ratios at the source, and their flavor ratios must be treated
separately. 
Note that this is a simplified scenario since 
there would be always some contamination of 
the $\bar{\nu}_e$ flux in a realistic situation, 
see \cite{Hummer:2010ai,Hummer:2010vx}.
However, we assume the ratios in eq. (\ref{eq:flavor-ratio-italic-pgamma})
for the $p\gamma$ source for the purpose of the illustration of the main point of this work.

Here again $\mu^+$'s decays can be damped by energy losses, in which
case no antineutrinos would be produced at all at the source, leading
to a even more asymmetric case.  
Therefore, as pointed out in \cite{Shoemaker:2015qul}, 
in general, to study $p\gamma$ sources, 
assumption of equal flavor ratios for neutrinos and
antineutrinos is not appropriate.

\subsection{\bf Effect of oscillation}
\label{subsec:effect-of-oscillation}

According to our current understanding,  neutrino flavor eigenstates and mass eigenstates are 
connected by a mixing matrix $U$, 
so as they propagate mass eigenstates acquire relative phases 
that generate the so-called neutrino flavor oscillations. 
By propagating from the astrophysical
 source, where they are produced, to the Earth, where they are detected, 
these relative phases are washed out, so neutrinos arrive as 
an incoherent mixture of mass eigenstates~\cite{Athar:2000yw}.  
We can thus write the flux flavor ratios at the Earth as 
\begin{eqnarray}
{\rm f}^{\,\oplus}_{\nu_\alpha} 
({\rm f}^{\,\oplus}_{\overline{\nu}_\alpha})
= \sum_{\beta=e,\mu,\tau} \sum_{j=1,2,3} 
\vert U_{\alpha j}\vert^2 \vert U_{\beta j}\vert^2 
\; {\rm f}^{\,\rm S}_{\nu_\beta} 
({\rm f}^{\,\rm S}_{\overline{\nu}_\beta})\,  \quad \quad 
(\alpha = e, \mu, \tau), 
\end{eqnarray}
for neutrinos (antineutrinos).

So if we start at the source with equal fluxes for neutrinos and
antineutrinos this equality will be preserved by the incoherent
mixture that will arrive at the Earth.  Conversely, if their original
fluxes are unequal, one should not expect to have equal neutrino and
antineutrino fluxes arriving on Earth.  This feature has been studied
in connection with the Glashow resonance as a way to probe production
models~\cite{Berezinskii_Gazizov_1977,Barger:2014iua,Anchordoqui:2004eb,Ahlers:2005sn,Bhattacharjee:2005nh,Pakvasa:2007dc,Xing:2011zm,Bhattacharya:2011qu} and beyond the Standard Model
physics~\cite{Shoemaker:2015qul}.  In the next section we discuss how
the difference in neutrino and antineutrino fluxes can affect the
fitting of IceCube showers and track events in a way to mimic new
physics.  Since we do not know the production mechanism, it is
imperative to fit the data without this limiting assumption, including
extra parameters to allow for the possibility of a difference in
neutrino and antineutrino fluxes for each flavor.

\section{On the Interpretation of IceCube Data}
\label{sec:pgamma}

We will show here that if the fluxes of high energy extraterrestrial
neutrinos and antineutrinos that arrive at the Earth were unequal, by
assuming their equality in the fit of IceCube data, one can
obtain allowed regions in the flavor-ratio
triangle at the Earth that lay outside the region expected by standard
neutrino oscillations.  This can be misinterpreted as a sign of new
physics, while in fact it is simply due to an incorrect assumption
about the fluxes. In practice, we are going to study four scenarios, which include two kinds of sources and two different 
exposures at IceCube. 

In the first part of each analysis we simulate the number of showers and tracks that IceCube would observe given a particular source and exposure. In order to do this we fix the set of neutrino fractions at the source and then we derive the fractions at the Earth by using standard values for the neutrino mixing matrix. Afterwards, we use this as input to predict the number of track and shower events for a given energy bin of the IceCube detector. In fact, we just use the fractions to scale up a set of normalized values of showers and tracks that we have already computed following the detailed prescriptions of Ref.~\cite{Palomares-Ruiz:2015mka} (see Appendix B for details). In these computations we have assumed that neutrinos and antineutrinos of any flavor $\alpha$ follow the same energy spectrum, so that we can parameterize the sum of the
fluxes as

\begin{equation}
\frac{d\Phi(E,\gamma,C_{r})}{dE} =  \frac{C_{r}}{10^{8}}\bigg(\frac{E}{100\, \text{TeV}}\bigg)^{2-\gamma}\frac{1}{E^{2}} \quad
[{\rm GeV^{-1} \, cm^{-2} \, str^{-1} \, s^{-1}}]\, ,
\label{eq:fluxatearth}
\end{equation}
\noindent
where $E$ is the neutrino energy, $C_r$ is a normalization constant and $\gamma$ is the spectral index. Thus, the neutrino (antineutrino) flavor fluxes at the Earth are
\begin{equation}
\frac{d\Phi_{\nu_\alpha(\overline{\nu}_\alpha)}^{\oplus}(E,\gamma,C_{r})}
{dE} = \text{f}_{\nu_\alpha(\overline{\nu}_\alpha)}^{\,\oplus} \;
\frac{d\Phi(E,\gamma,C_{r})}{dE}\, ,
\end{equation} 

\noindent where $\text{f}_{\nu_\alpha(\overline{\nu}_\alpha)}^{\,\oplus}$ are the normalized neutrino fractions at the Earth. For each simulation of the expected data we fix the spectral index to $\gamma=2.5$ and the total number of events of astrophysical neutrinos to $N_{a}=18\times 10\ (100)$ in the energy interval [28 TeV, 10 PeV]~. These numbers are motivated by the best fit point of IceCube given by the combined maximum-likelihood analysis of Ref.~\cite{Aartsen:2015knd}, 
which corresponds to 18 astrophysical neutrinos in 988 days in the energy range [28 TeV, 10 PeV], so we use 10 (100) times more exposure to try to simulate future observations. 

Furthermore, we also consider the background from atmospheric neutrinos
and atmospheric muons. Again, we mostly follow
Ref.~\cite{Palomares-Ruiz:2015mka} (see Appendix C for details). 
Thus, in each analysis we fix the number of atmospheric neutrinos to
$N_{\nu}=6.6$ and the number of atmospheric muons to $N_{\mu}=8.4$
per 18 astrophysical neutrino events.\footnote{Indeed, we fix
these numbers to $N_{\nu}=6.6\times 10\ (100)$ and $N_{\mu}=8.4\times
10\ (100)$ in order to correct for the two exposures that we are
simulating. }
This was done in order to match the central values obtained by IceCube in the interval $[28\,\text{TeV,} 10\,\text{PeV]}$ after 988 days of data taking \cite{Aartsen:2014gkd}. The uncertainty associated to these numbers are not considered in this analysis but we have checked that this effect, that may be added through nuisance parameters in the fit, does not change our final conclusions. 

Besides, it would be necessary to take into account the potential effects associated to 
the experimental misidentification of tracks as showers,
as done in Ref.~\cite{Palomares-Ruiz:2015mka}
where the rates of 20 and 30\% were considered based on \cite{Aartsen:2015ivb}.
In this work, we set the track to shower miss-identificatin rate to be 
somewhat more optimistic value, $10\%$, 
which is applied to every contribution. 
We believe that moderate modifications
of this rate do not affect essentially our final conclusions. 
The final number of showers and tracks computed in this way will simulate the observed data measured by IceCube. Finally, let us emphasize that we consider only one realization of the simulation for each scenario (see Appendix A for more details).

In the second part of each analysis we study the fit of the simulated data. Thus, we use the same approach described above to compute the number of showers and tracks, but now considering floating values of the normalization parameter $N_a$, the spectral index $\gamma$ and the neutrino fractions at the Earth f$^{\oplus}_{\nu_\alpha(\overline{\nu}_\alpha)}$. For each scanned point in this parameter space we evaluate a binned likelihood function, that we extreme in order to find the best fit point and the corresponding confidence level regions of the relevant parameters. In the fit procedure we make the key assumption that the neutrino and antineutrino fluxes are equal for all flavors, {\em i.e.}, f$^{\,\oplus}_{\nu_\alpha}$= f$^{\,\oplus}_{\overline{\nu}_\alpha}$= $f^{\oplus}_{\nu_\alpha}/2$. Details about the fit procedure can be found in Appendix A.

\subsection{A Possible Scenario Fitted with the Wrong Assumption}

We start by  discussing a theoretically viable example.
Let's assume here that the neutrino  fluxes are  
produced at the source by the mechanism $p\gamma$ of photo-disintegration.
This corresponds to fixing 
the neutrino and antineutrino flavor fractions at the source as 

\begin{eqnarray}
({\rm f}^{\,\rm S}_{\nu_{e}}: {\rm f}^{\,\rm S}_{\nu_{\mu}}: 
{\rm f}^{\,\rm S}_{\nu_{\tau}})
= \left(\frac{1}{3} : \frac{1}{3} : 0\right)_{{\rm S}}, \quad \quad 
({\rm f}^{\,\rm S}_{\overline{\nu}_{e}}: {\rm f}^{\,\rm S}_{\overline{\nu}_{\mu}}:
{\rm f}^{\,\rm S}_{\overline{\nu}_{\tau}})
= \left(0 : \frac{1}{3} : 0\right)_{\rm S}.
\end{eqnarray}
For definiteness, we also fix the oscillation parameters to 
$$\sin^2\theta_{12}=0.31 \, , \; \sin^{2}\theta_{23}=0.60\, , \; 
\sin^2\theta_{13}=0.02 \, , \; \delta=3\pi/2, $$ 
in order to obtain the theoretical predicted 
propagated flavor fractions at the Earth, 
\begin{eqnarray}
\hskip -1cm
({\rm  f}_{\nu_{e}}^{\oplus \rm T}: {\rm f}_{\nu_{\mu}}^{\oplus \rm T}:
 {\rm f}_{\nu_{\tau}}^{\oplus \rm T}) &= &(0.25 : 0.21 : 0.21)_\oplus, \ \ 
({\rm  f}_{\overline{\nu}_{e}}^{\oplus \rm T}: 
{\rm  f}_{\overline{\nu}_{\mu}}^{\oplus \rm T}: 
{\rm  f}_{\overline{\nu}_{\tau}}^{\oplus \rm T})=(0.06 : 0.15 : 0.12)_{\oplus},
\label{eq:flavor-ratio-earth-scenario1}
\end{eqnarray}
where the superscript T here means theoretically expected value.
Note that, as expected, the sum of the neutrino (antineutrino) fractions at the Earth
coincide with the sum of the neutrino (antineutrino) fractions at the source, and
similarly for antineutrinos. Also  
${\rm f}_{\nu_{\alpha}}^{\,\oplus \rm T}+
{\rm  f}_{\overline{\nu}_{\alpha}}^{\,\oplus \rm T}\approx 1/3$, 
albeit ${\rm  f}_{\nu_{\alpha}}^{\,\oplus \rm T} \neq 
{\rm  f}_{\overline{\nu}_{\alpha}}^{\,\oplus \rm T}$. As we said before, these fractions
fix the {\em mock-data} for the fit. Our binned maximum-likelihood fit (see Appendix A) of this data 
involves four free parameters: the global normalization,
$N_{a}^{\rm F}$, the spectral index, $\gamma^{\rm F}$, and the
$\nu_e+\bar{\nu}_e$ and $\nu_\mu+\bar{\nu}_\mu$ flavor fractions at
the Earth $f^{\oplus \rm F}_{\nu_{e}}, f^{\oplus \rm F}_{\nu_{\mu}}$
with the constraint $f^{\oplus \rm F}_{\nu_{\tau}}=1-f^{\oplus \rm
  F}_{\nu_{e}}-f^{\oplus \rm F}_{\nu_{\mu}}$.  The superscript F here
means fit value.  As the result of the fit we get the best fit point:
$N_{a}^{\rm F}=17.91\times 10 (100)$, $\gamma^{\rm F}=2.48$,
$f^{\oplus\rm F}_{\nu_{e}}=0.098$ and $f^{\oplus \rm
  F}_{\nu_{\mu}}=0.27$.

\begin{figure}[htb]
\vglue -1.1cm
\centering
\begin{minipage}{1.0\textwidth}
\begin{minipage}{.5\textwidth}
\centering
\includegraphics[width=1.\linewidth]{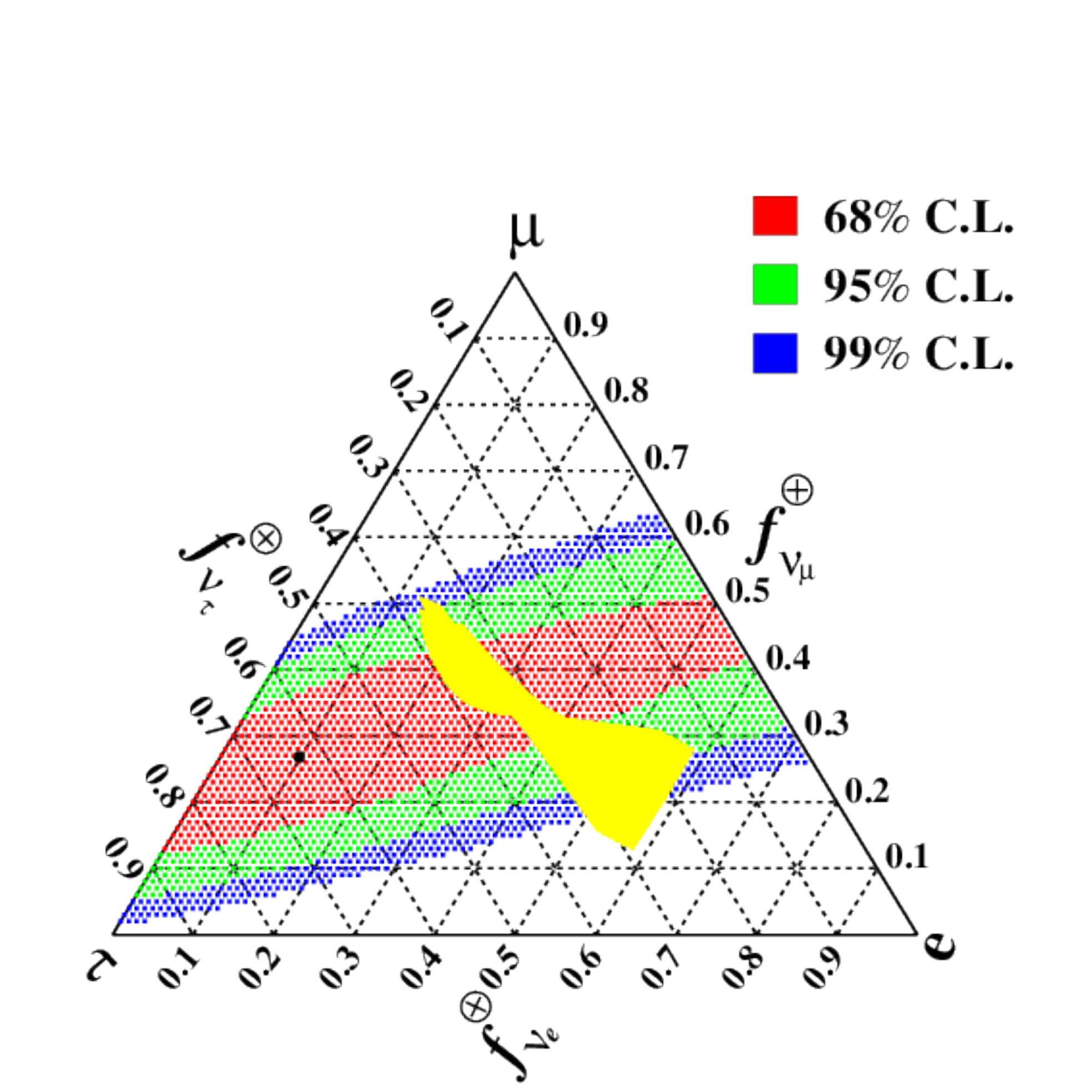}
\end{minipage}
\begin{minipage}{.5\textwidth}
\centering
\includegraphics[width=1.\linewidth]{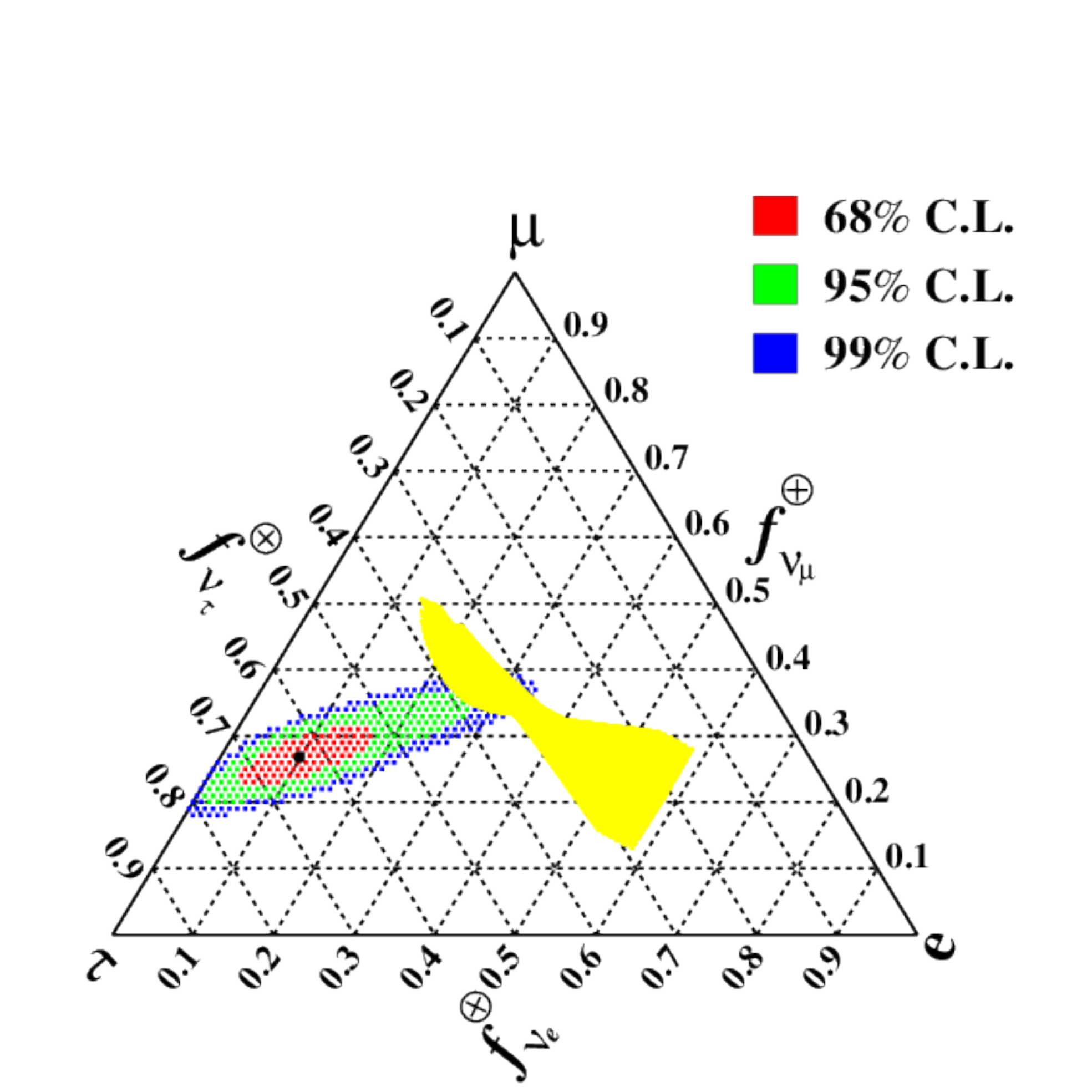}
\end{minipage}
\end{minipage}
\vglue 0.2cm
\caption{\label{fig2:triangle-allowed-regions-ex1} Regions allowed at
  68\% (red), 95\% (green) and 99\% CL (blue) in the parameter space of
  the astrophysical flavor ratios $(f^\oplus_{\nu_{e}},
  f^\oplus_{\nu_{\mu}},f^\oplus_{\nu_{\tau}})$ at the Earth.
On the left (right) panel we used a total number of
  180 (1800) astrophysical neutrino events. The input data was
  simulated for the $p\gamma$ production scenario with 
$({\rm f}^{\,\rm S}_{\nu_{e}}: {\rm f}^{\,\rm S}_{\nu_{\mu}}: 
{\rm  f}^{\,\rm S}_{\nu_{\tau}})=(\frac{1}{3}:\frac{1}{3}:0)_{\rm S}, 
({\rm f}^{\,\rm S}_{\overline{\nu}_{e}}: 
{\rm  f}^{\,\rm S}_{\overline{\nu}_{\mu}}: 
{\rm  f}^{\,\rm S}_{\overline{\nu}_{\tau}})=(0:\frac{1}{3}:0)_{{\rm S}}$ 
and  $\sin^2\theta_{12}=0.31 \, , \sin^{2}\theta_{23}=0.60\, ,
\sin^2\theta_{13}=0.02 \, , \delta=3\pi/2$ and the data was fitted
with the wrong assumption that all neutrino and antineutrino flavor
fluxes are the same at the Earth.  We also show the region allowed
by the standard oscillation hypothesis (same as Fig.~\ref{fig1:triangle-standard-expected}).}
\end{figure}

We show the results of our fit in Fig.~\ref{fig2:triangle-allowed-regions-ex1}
for the cases where the total number of astrophysical neutrino events 
are assumed to be 180 (left panel) and 1800 (right panel).
It is interesting to notice that, with the wrong assumption about neutrino and antineutrino fluxes,
to fit the {\em mock-data} corresponding to the $p\gamma$ scenario 
requires that the best fit point for the flux flavor fractions falls outside of the standard
central region of the flavor-ratio triangle, as can be clearly seen
by the solid black circles in 
Fig.~\ref{fig2:triangle-allowed-regions-ex1}. In this figure we show
the allowed regions at 68\% (red), 95\% (green) and 99\% CL (blue) in 
the flavor-ratio triangle spanned in the 
$(f^\oplus_{\nu_{e}}, f^\oplus_{\nu_{\mu}},f^\oplus_{\nu_{\tau}})$
parameter space. 
The limited region
predicted by standard oscillations, as shown before in
Fig.~\ref{fig1:triangle-standard-expected}, is also depicted in yellow.
In the left panel, we note that despite that the wrong assumption
(of equal fluxes of neutrino and antineutrinos) was made,
the result of the fit is quite good such that 
this assumption can not be statistically rejected even at $1\sigma$ confidence level. However, in the
right panel, when ten times more statistics is used, just one small part of the yellow zone is compatible with the boundary of the $2\sigma$ confidence level region.

How can we understand this result?  Let's see how it arises from the
fact that the simulated data, producing flavor fractions at the Earth
$\text{f}_{\nu_\alpha}^{\,\oplus \rm T}\neq
\text{f}_{\bar{\nu}_\alpha}^{\,\oplus \rm T}$,  
is fitted using
$\text{f}_{\nu_\alpha}^{\,\oplus \rm  F} = 
\text{f}_{\bar{\nu}_\alpha}^{\, \oplus \rm F} = 
f_{\nu_\alpha}^{\oplus  F}/2$ at the Earth. 
The argument goes as follow.  The spectrum of
shower events from $\bar{\nu}_{e}$, thanks to the Glashow Resonance,
is quite different from the spectra of shower events from $\nu_{e}$,
$\nu_{\tau}$ and $\bar{\nu}_{\tau}$ 
(see Fig.~\ref{fig6:Signal-contribution-to-ShTr}). With
enough statistics, as it is the case of our simulated events, we
expect the fitted fractions to reproduce theory fractions. 
Although the Glashow Resonance is localized around one or two energy
bins we note that its effect is going to affect the fit of 
the whole energy range because it determines the overall
normalization of $\bar{\nu}_e$ flux.
Later we can check this hypothesis
from the comparison between the fit and the expected fractions. So the
$\bar{\nu}_e$ shower event distribution will impose

\begin{equation}
\frac{f_{\nu_{e}}^{\oplus \rm F}}{2}   \approx  
\text{f}_{\bar{\nu}_{e}}^{\,\oplus \rm T} \simeq 0.06\, ,
\label{eq:fe}
\end{equation}
while  since $\nu_{e}$, $\nu_{\tau}$ and
$\bar{\nu}_{\tau}$ shower events have similar 
energy dependence (see Fig.~\ref{fig6:Signal-contribution-to-ShTr})
they impose 

\begin{equation}
\frac{f_{\nu_{e}}^{\oplus \rm F}}{2}+f_{\nu_{\tau}}^{\oplus \rm F}  
\approx  \text{f}_{\nu_{e}}^{\,\oplus \rm T} + 
\text{f}_{\nu_{\tau}}^{\,\oplus \rm T}+\text{f}_{\bar{\nu}_{\tau}}^{\,\oplus \rm T}\, ,
\label{eq:fs}
\end{equation}
and track events will fix 
\begin{equation}
f_{\nu_{\mu}}^{\oplus \rm F}   \approx  
\text{f}_{\nu_{\mu}}^{\,\oplus \rm T}  + 
\text{f}_{\bar{\nu}_{\mu}}^{\,\oplus \rm T} 
\simeq 0.36 \, .
\label{eq:fmu}
\end{equation}

By combining equations (\ref{eq:fe}) and (\ref{eq:fs}), 
and $\text{f}_{\nu_{e}}^{\,\oplus \rm T}\simeq 0.25$ 
from eq.~(\ref{eq:flavor-ratio-earth-scenario1}),
we get
\begin{eqnarray}
f_{\nu_{\tau}}^{\oplus \rm F} & \approx & 
\text{f}_{\nu_{\tau}}^{\,\oplus \rm T}+\text{f}_{\bar{\nu}_{\tau}}^{\,\oplus\rm T}
+\text{f}_{\nu_{e}}^{\,\oplus \rm T}-\text{f}_{\bar{\nu}_{e}}^{\,\oplus
\rm T},\nonumber \\
& \approx & \frac{1}{3}+
\text{f}_{\nu_{e}}^{\,\oplus \rm T}-\text{f}_{\bar{\nu}_{e}}^{\,\oplus \rm T}
\simeq 0.52\, ,
\end{eqnarray}
where we have used that 
$\text{f}_{\nu_{\tau}}^{\,\oplus \rm  T}+
\text{f}_{\bar{\nu}_{\tau}}^{\,\oplus \rm T}\sim 1/3$.  
We see that the output of the fit is consistent with the rough estimation given by
eqs.~(\ref{eq:fe})--(\ref{eq:fmu}) based on our event distribution
considerations for showers of different flavors and the importance of
the Glashow Resonance for the fit. Notice that the previous equations do not 
include the relative weight of the distributions, we have checked that by considering this extra 
information the agreement is even better. 

The conclusions derived from this case will certainly hold for a variety of other examples. For instance, if no antineutrinos at all were produced at the source, and consequently none would arrive at the Earth (the case of $\mu^+$ damping), we would expect to fit $f_{\nu_{e}}^{\oplus \rm F} \approx \text{f}_{\bar{\nu}_{e}}^{\,\oplus  \rm T} \sim 0$, leading to a best fit point again well outside the standard region of the flavor-ratio triangle.

\subsection{Re-thinking Current Results}

In Ref.~\cite{Aartsen:2015knd} IceCube finds, assuming
$\text{f}_{\nu_\alpha}^{\,\oplus}=\text{f}_{\bar{\nu}_\alpha}^{\,\oplus}
=f_{\nu_\alpha}^{\oplus}/2$,
the best fit point for the neutrino flavor ratios at the Earth to be
$(f^\oplus_{\nu_{e}}: f^\oplus_{\nu_{\mu}}:f^\oplus_{\nu_{\tau}})
\simeq (0.5:0.5:0.0)$. 
While it is not yet statistically significant, 
this point is outside the {\em standard region} of
Fig.\ref{fig1:triangle-standard-expected}. 
Let's assume, for the sake of discussion, that these
numbers will be confirmed with higher significance
by future data. 
Is this a sign of exotic physics (due to some non-standard properties of neutrinos)
or can this be explain by standard physics?  Can we find
neutrino and antineutrino flavor ratios at the source that can give
this result at the Earth if fitted with the wrong assumption?

Indeed, we can explain the IceCube best fit of neutrino fractions by
considering only standard neutrino physics plus an asymmetric source of
neutrinos. 
Take, for instance, the following flavor fractions at the source 
\begin{eqnarray}
({\rm f}^{\rm S}_{\nu_{e}}: {\rm f}^{\rm S}_{\nu_{\mu}}: 
{\rm f}^{\rm S}_{\nu_{\tau}})=(0 : 0.3 : 0)_{\rm S}, \ \ 
({\rm f}^{\,\rm S}_{\overline{\nu}_{e}}: 
{\rm  f}^{\,\rm S}_{\overline{\nu}_{\mu}}: 
{\rm  f}^{\,\rm S}_{\overline{\nu}_{\tau}})=(0.4:0.3:0)_{\rm S}.
\label{eq:flavor-ratio-example2}
\end{eqnarray}

We note that if we do not distinguish neutrinos and antineutrinos, 
the flavor ratios given in (\ref{eq:flavor-ratio-example2})
correspond to 
$(f^{\rm S}_{\nu_{e}}: f^{\rm S}_{\nu_{\mu}}:f^{\rm S}_{\nu_{\tau}})
= (0.4:0.3:0.3)$ which gives
$(f^\oplus_{\nu_{e}}: f^\oplus_{\nu_{\mu}}:f^\oplus_{\nu_{\tau}})
\simeq (0.34:0.34:0.32)$, very close to the central point of the triangle. 
Below we will see that this point could be mistakenly excluded due to 
the wrong assumption provided that we will have enough statistics,
giving (fake) best fit outside the standard allowed region.

Note that this scenario is in principle possible if we consider the case where 
neutrinos are coming from the two kinds of sources, namely the combination of the pion decay with 
muon dump in $pp$ scenario and neutron decay source with the ratios of the former to the latter 
equal to 4:3. In such a case, the $pp$ source leads to equal fluxes of $\nu_{\mu}$ and $\bar{\nu}_{\mu}$ 
whereas the neutron source leads to pure $\bar{\nu}_{e}$ flux. Now, fix
the same oscillation parameters as before to propagate this to the
Earth. The result at the Earth will be 
\begin{eqnarray}
\hspace{-0.5cm}
({\rm f}^{\,\oplus \rm T}_{\nu_{e}}: {\rm f}^{\,\oplus \rm T}_{\nu_{\mu}}: 
{\rm f}^{\,\oplus  \rm T}_{\nu_{\tau}}) = (0.06 : 0.13 : 0.11)_{\oplus}, \ \ 
({\rm f}^{\,\oplus \rm  T}_{\overline{\nu}_{e}}: 
{\rm f}^{\,\oplus \rm  T}_{\overline{\nu}_{\mu}}: 
{\rm f}^{\,\oplus \rm  T}_{\overline{\nu}_{\tau}})=(0.28:0.21:0.21)_{\oplus}.
\end{eqnarray}
We have generated {\em mock data} in accordance with this theoretical
prediction at the Earth assuming $N_{a}^{\rm T}=18 \times 10\, (100)$
astrophysical neutrinos and $\gamma^{\rm T}=2.5$.  
Finally, we have fit the expected number of events using the standard
approach of IceCube, {\em i.e.}, assuming the neutrino and antineutrino fractions
to be equal at the Earth.  

\begin{figure}[th!]
 \vglue -1.0cm
\includegraphics[width=0.5\linewidth]{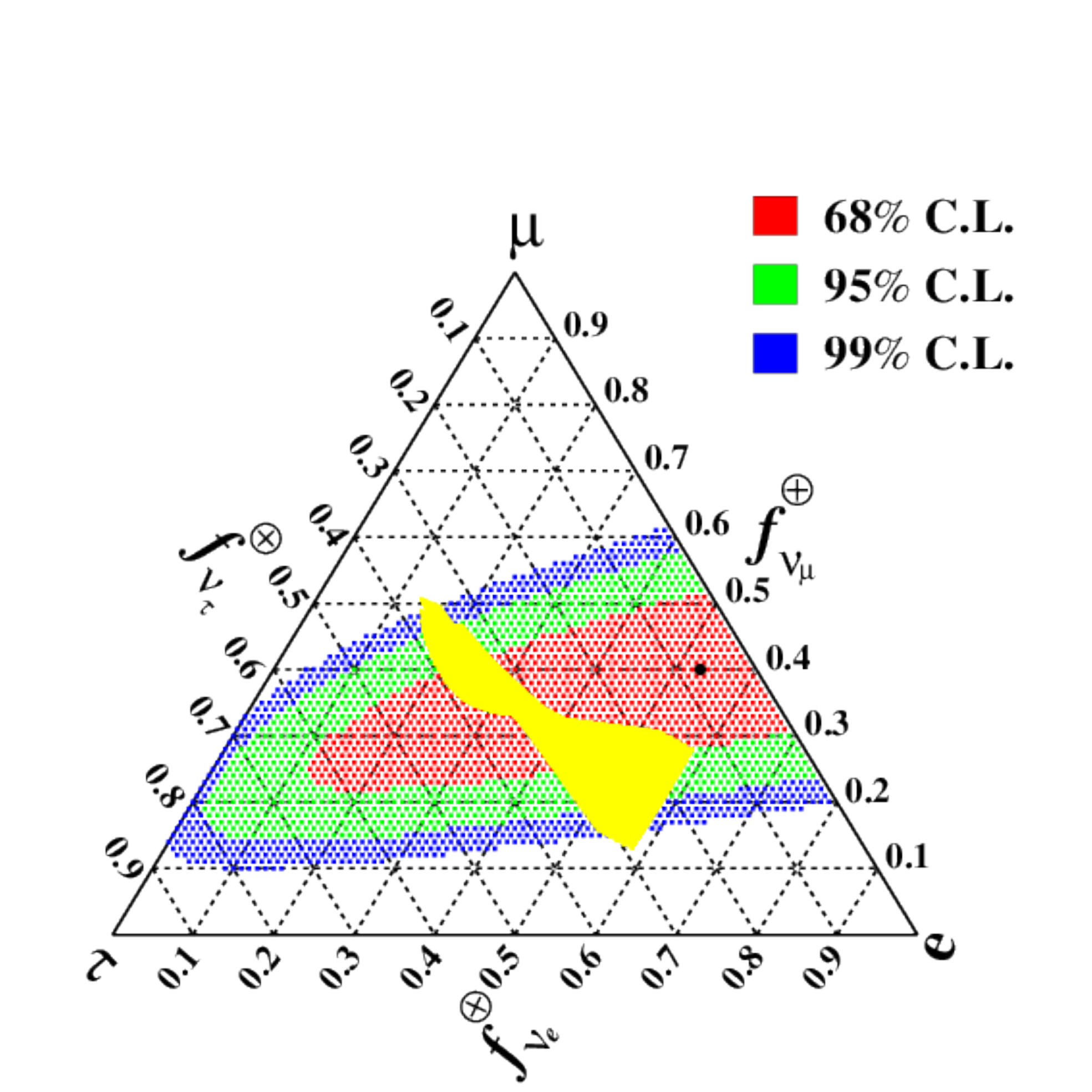}
\includegraphics[width=0.5\linewidth]{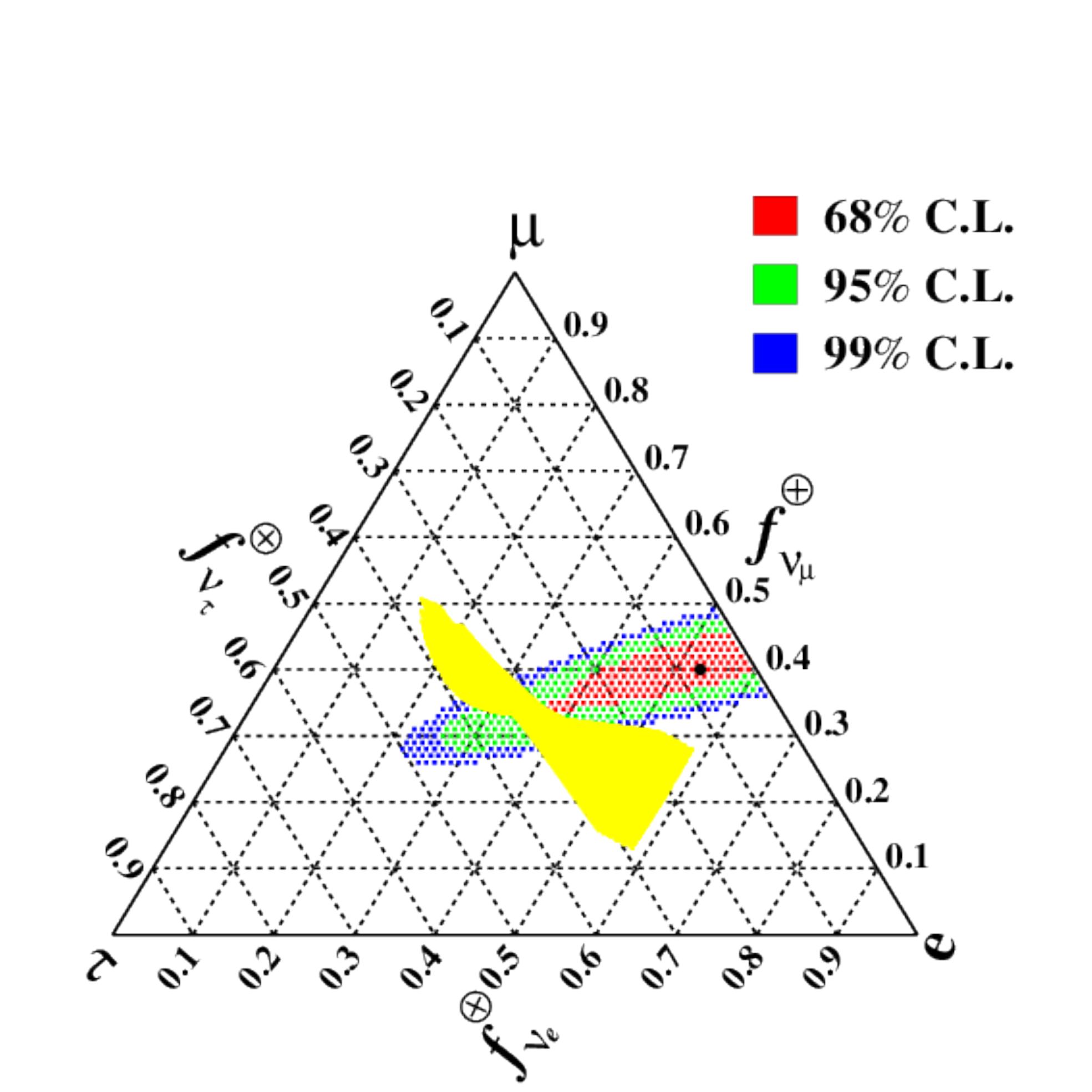}
\vglue 0.2cm
\caption{\label{fig3:triangle-allowed-regions-ex2} 
Allowed regions at 68\% (red), 95\% (green) and
  99\% CL (blue) for the astrophysical flavor ratios
  $(f^\oplus_{\nu_{e}}, f^\oplus_{\nu_{\mu}},f^\oplus_{\nu_{\tau}})$
  at the Earth. On the left (right) panel we used a total number of
  180 (1800) astrophysical neutrino events.
The input data was simulated for a fictitious
  production scenario with 
$({\rm f}^{\,\rm S}_{\nu_{e}}: {\rm f}^{\,\rm S}_{\nu_{\mu}}: 
{\rm f}^{\,\rm S}_{\nu_{\tau}})=(0:0.3:0)_{{S}}, 
({\rm  f}^{\,\rm S}_{\overline{\nu}_{e}}: 
{\rm   f}^{\,\rm S}_{\overline{\nu}_{\mu}}: 
{\rm   f}^{\,\rm S}_{\overline{\nu}_{\tau}})=(0.3:0.4:0)_{{S}}$, 
and  $\sin^2\theta_{12}=0.31 \, , \sin^{2}\theta_{23}=0.60\, ,
\sin^2\theta_{13}=0.02 \, , \delta=3\pi/2$. The data was fitted with
the wrong assumption that all neutrino and antineutrino flavor
fluxes are the same at the Earth. We also show the region 
predicted by the standard oscillation hypothesis 
(same as in Fig.~\ref{fig1:triangle-standard-expected}).}
\end{figure}

We show our results in Fig.~\ref{fig3:triangle-allowed-regions-ex2}.
The result of our fit, with this wrong
assumption, is $(f^{\oplus \rm F}_{\nu_{e}}: f^{\oplus \rm
  F}_{\nu_{\mu}}:f^{\oplus \rm F}_{\nu_{\tau}})=(0.52:0.41:0.07)$,
$N_{a}^{\rm F}=179$ and $\gamma^{\rm F}=2.46$. These results are in
close agreement with the IceCube fit, as shown in
Fig. \ref{fig3:triangle-allowed-regions-ex2}. Furthermore, this is consistent with
eqs.~(\ref{eq:fe})--(\ref{eq:fmu}).  We conclude that there could be a
standard physics explanation for this point as long as we relax the
constraint that imposes flavor fraction to be the same for neutrinos
and antineutrinos.  
We have checked explicitly that if we perform a fit by varying 
independently neutrino and antineutrino fluxes, 
we recover the input fractions and the allowed regions 
sit around the central part of the flavor-ratio triangle.
%
%

\begin{figure}[h!]
\centering
\begin{minipage}{1.0\textwidth}
\begin{minipage}{.5\textwidth}
\centering
\includegraphics[width=\linewidth]{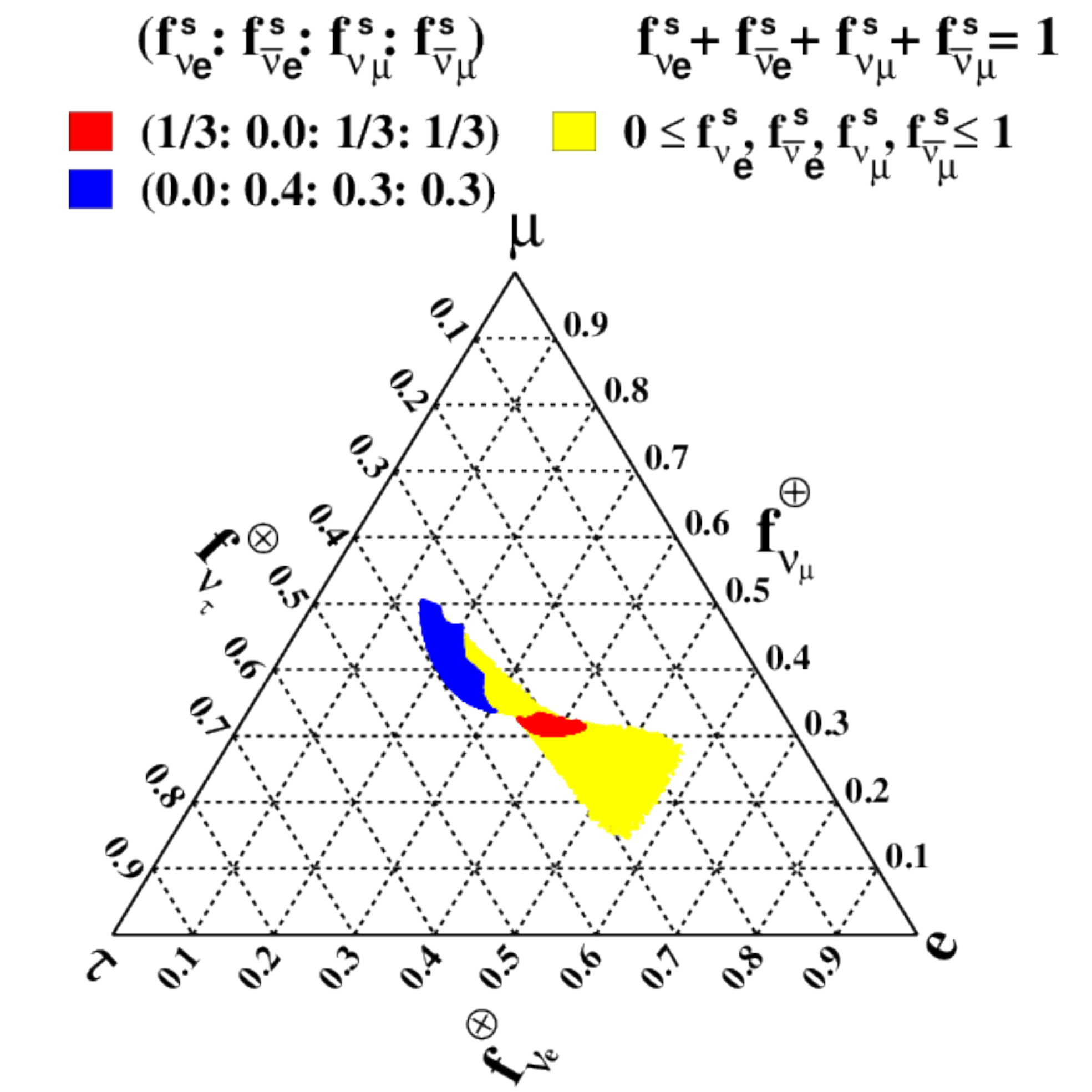}
\end{minipage}
\begin{minipage}{.5\textwidth}
\centering
\includegraphics[width=\linewidth]{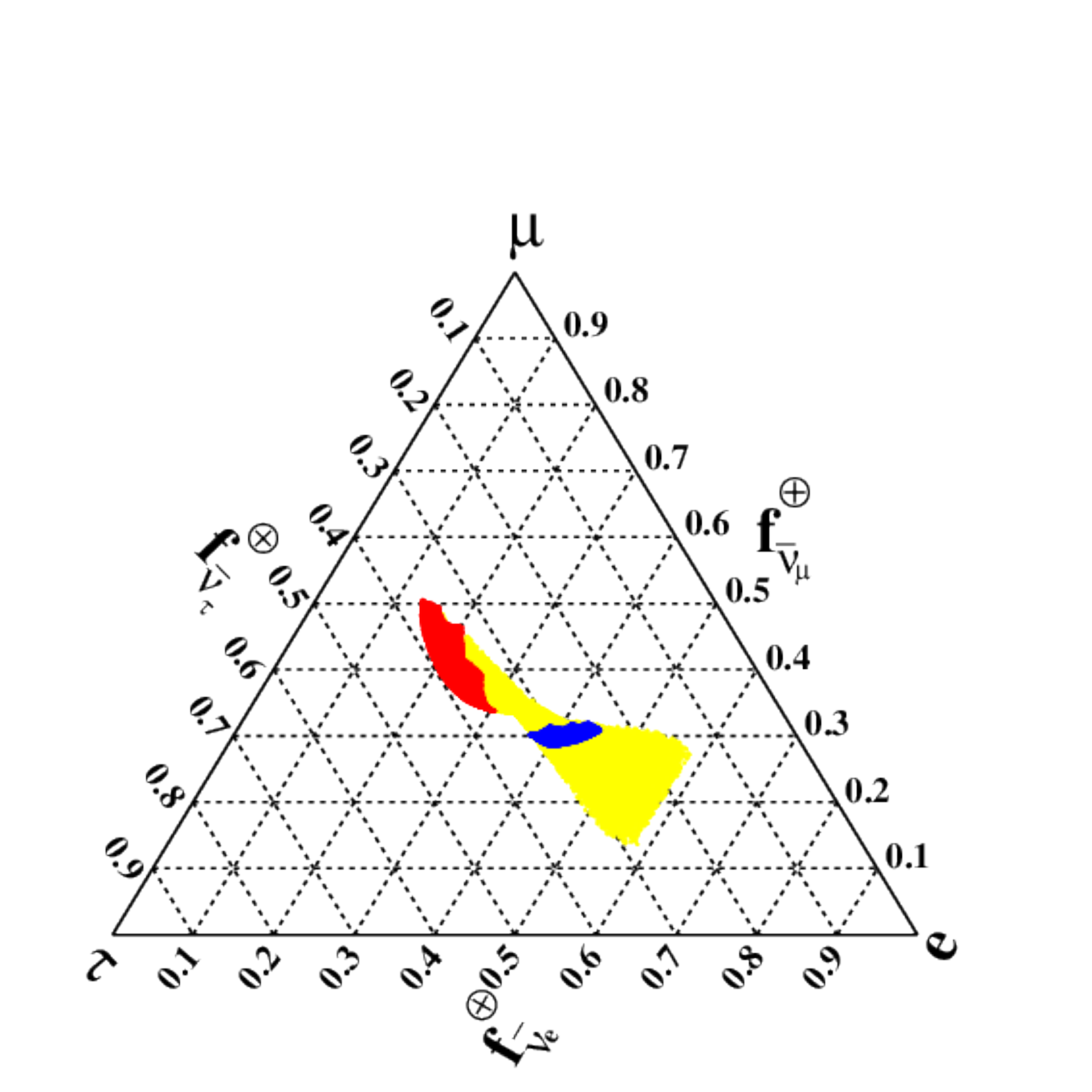}
\end{minipage}
\end{minipage}
\caption{\label{fig4:triangle-separate} On the right (left) panel we
  present the currently allowed regions for the flux flavor ratios
  f$^{\,\oplus}_{\nu_e}$, f$^{\,\oplus}_{\nu_\mu}$,
  f$^{\,\oplus}_{\nu_\tau}$ (f$^{\,\oplus}_{\bar{\nu}_e}$,
  f$^{\,\oplus}_{\bar{\nu}_\mu}$, f$^{\,\oplus}_{\bar{\nu}_\tau}$) at
  the Earth by various high energy neutrino production mechanisms .
  We let neutrino mixing parameters vary with the current 3$\sigma$
  allowed regions~\cite{Bergstrom:2015rba}.  We have normalized the
  total neutrino (left panel) and antineutrino (right panel) flavor
  fractions at the Earth to 1 separately for each point in order to
  show them in this manner.  The red regions correspond to the case
  where (f$^{\,\rm S}_{\nu_e}$, f$^{\,\rm S}_{\nu_\mu}$, f$^{\,\rm
    S}_{\nu_\tau}$) = (1/3 : 1/3 : 0) and (f$^{\,\rm
    S}_{\bar{\nu}_e}$, f$^{\,\rm S}_{\bar{\nu}_\mu}$, f$^{\,\rm
    S}_{\bar{\nu}_\tau}$)=(0 : 1/3 : 0) and the blue one corresponds
  to (f$^{\,\rm S}_{\nu_e}$, f$^{\,\rm S}_{\nu_\mu}$, f$^{\,\rm
    S}_{\nu_\tau}$)=(0 : 0.3 : 0) and (f$^{\,\rm S}_{\bar{\nu}_e}$,
  f$^{\,\rm S}_{\bar{\nu}_\mu}$, f$^{\,\rm S}_{\bar{\nu}_\tau}$)=(0.4
  : 0.3 : 0).}
\end{figure}

\begin{figure}[htb]
\centering
\includegraphics[width=0.5\linewidth]{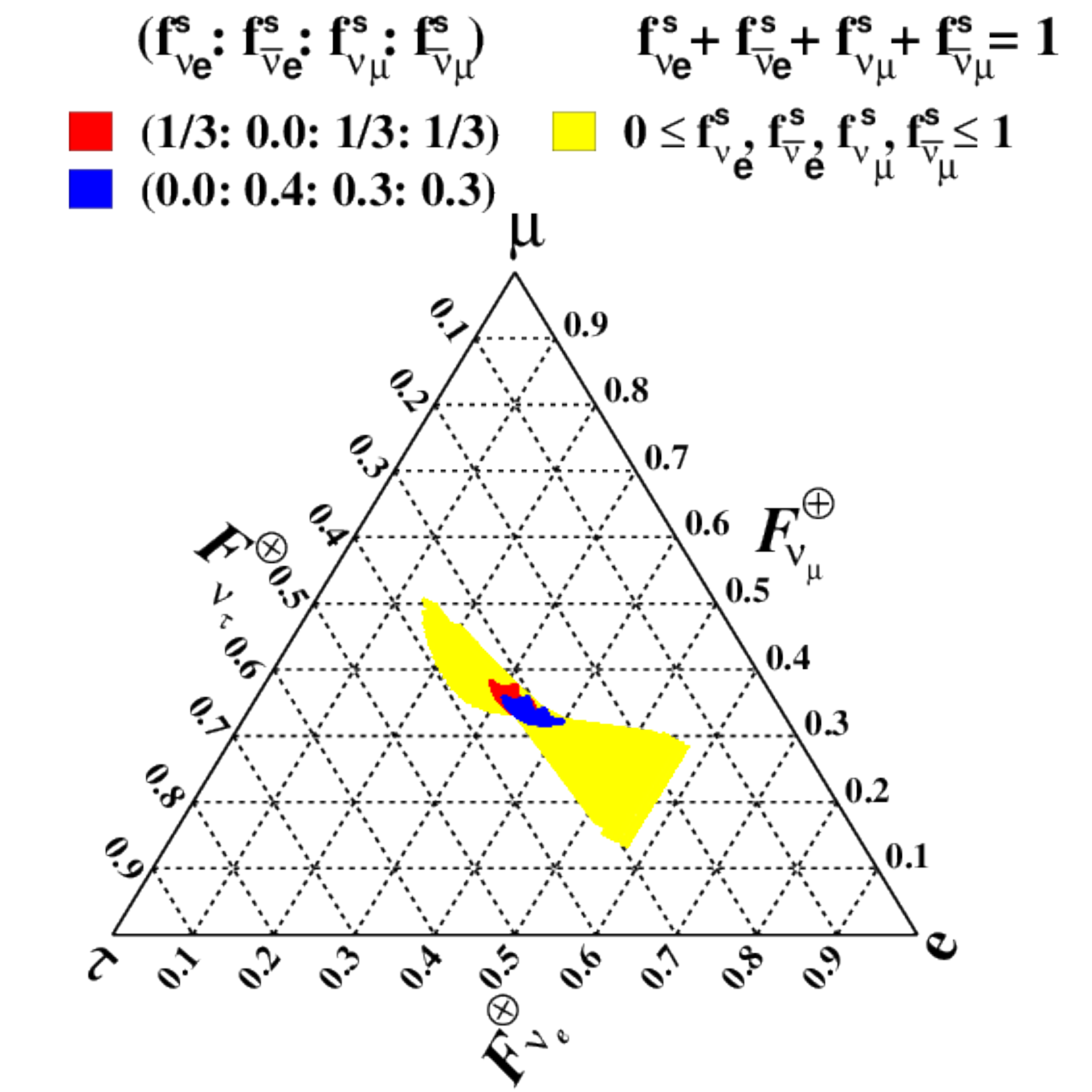}
\caption{\label{fig5:triangle-sum} 
Same as Fig.~\ref{fig4:triangle-separate} 
but for the sum of the
  neutrino and antineutrino fractions,
  $\mathbf{F^\oplus_{\nu_\alpha}=f^\oplus_{\nu_\alpha}+f^\oplus_{\bar{\nu}_\alpha}}$.}
\end{figure}
%

\section{Discussions and Conclusions}
\label{sec:conclusions}
Whether $pp$ or $p\gamma$ interactions (or some combination) in Fermi
engines are the sources of these neutrinos will be probably
established by data, as soon as enough statistics is accumulated by
IceCube or its successors IceCube-Gen2~\cite{::2015dwa} or
KM3NeT2.0~\cite{Adrian-Martinez:2016fdl}.  Many
authors~\cite{Barger:2014iua,Shoemaker:2015qul,Berezinskii_Gazizov_1977,Anchordoqui:2004eb,Ahlers:2005sn,Bhattacharjee:2005nh,Pakvasa:2007dc,Xing:2011zm,Bhattacharya:2011qu} have
stressed the importance of the Glashow resonance as a discriminator.
 
In this paper we show that one must be careful when discussing IceCube
flavor ratio results, that at the moment assume neutrino and
antineutrino flavor ratios at the Earth are equal, as a way to probe
new physics in the neutrino sector.  The flavor ratios for neutrinos
and antineutrinos can be quite different as we do not know how they
are produced in the astrophysical environment, 
so the flavor ratios
for neutrinos and antineutrinos at the Earth also can be quite distinct, 
as can be seen in Fig.~\ref{fig4:triangle-separate}
where the expected normalized ratios for neutrinos (left panel)
and antineutrinos (right panel) are
shown separately for the two representative cases 
we considered in this work, 
$({\rm f}^{\,\rm S}_{\nu_{e}}: {\rm f}^{\,\rm S}_{\overline{\nu}_{e}}: 
{\rm f}^{\,\rm S}_{\nu_{\mu}}: {\rm f}^{\,\rm S}_{\overline{\nu}_{\mu}})
= 
(1/3 : 0: 1/3: 1/3)_{\rm S}$ in red
and $(0 : 0.4 : 0.3: 0.3)_{\rm S}$ in blue. Since in Fig.~\ref{fig4:triangle-separate} we only show normalized ratios, the information about the total fraction of neutrinos (antineutrinos) for each point is not present. A possible way of showing the neutrino and antineutrino ratios without losing information is by using 15 triangle plots with axes given by triplets (f$_{\nu_i}$, f$_{\nu_j}$, $1-$f$_{\nu_i}-$f$_{\nu_j}$) where $\nu_i$ runs over neutrino and antineutrino flavors\footnote{Indeed,  we may consider that effectively we have only four interesting ratios f$_{\nu_e}$, f$_{\bar{\nu}_e}$, f$_{\nu_\mu}$ and f$_{\nu_\tau}$, which reduces the number of triangle plots to 6. This because the spectra of muon and tau neutrinos are quite similar to their anti-neutrino counterpart.}.

In particular, we see that the standard case for $p\gamma$
production (in red) or the example we provided that could be
consistent with the current best fit point of IceCube (in blue), 
generate neutrino and antineutrino (normalized) ratios that cover different zones in the
respective flavor-ratio triangles.

On the other hand, the sum of the neutrino and antineutrino flavor
ratios at the Earth spread over a small region around the center of
the triangle, see Fig.~\ref{fig5:triangle-sum}.  However, when we fit
the data generated by these scenarios assuming equality between
neutrino and antineutrino ratios at the Earth, we obtain that the best
fit point lies outside the central region of the respective triangle
plot, as we showed in Figs.~\ref{fig2:triangle-allowed-regions-ex1}
and \ref{fig3:triangle-allowed-regions-ex2}.  Thus, this wrong
assumption could lead to the conclusion that new physics is necessary
in the neutrino sector, which is not necessarily the case.

Although at the current exposure of IceCube the assumption that
neutrino and antineutrino flavor fractions are the same at the Earth
when fitting the data seems quite reasonable, because most of the
triangle is still allowed by the current data (including the standard
region), this will have to change in the future when more data will
be available.  In particular, in order to be able to disentangle
different production mechanisms at the source from new physics in the
neutrino sector, it is imperative to work with the measured neutrino
and antineutrino flavor ratios at the Earth, separately.

\begin{acknowledgments}
We thank Olga Mena and Sergio Palomares-Ruiz for useful discussions on 
the computation of number of events induced by high energy astrophysical
neutrinos. We also acknowledge Kohta Murase and Walter Winter for useful
comments. This work
  was supported by Funda\c{c}\~ao de Amparo \`a Pesquisa do Estado de
  S\~ao Paulo (FAPESP), Funda\c{c}\~ao Carlos Chagas Fillho de
    Amparo \`a Pesquisa do Estado do Rio de Janeiro (FAPERJ) and
  Conselho Nacional de Ci\^encia e Tecnologia (CNPq).  H.N. and
  R.Z.F. thank the Kavli Institute for Theoretical Physics in UC Santa
  Barbara for its hospitality, where part of this work was initiated.
This research was supported in part by the National Science Foundation 
under Grant No. NSF PHY11-25915, and 
European Union's Horizon 2020 research and innovation programme under 
Marie Sklodowska-Curie grant agreement No 674896.

\end{acknowledgments}

\appendix
\section{Binned likelihood}

In order to compute both the best fit point plus the confidence level regions for 
the relevant free parameters, we consider as test statistics the function \cite{Agashe:2014kda},
\begin{equation}
-2\ln \lambda(\theta)  = 2\sum_{k=\rm sh,\rm tr}\sum_{i=1}^{\rm N_{\rm bins}} \bigg[ \mu_{k,i}(\theta) - n_{k,i} + n_{k,i}\ln\frac{n_{k,i}}{\mu_{k,i}(\theta)}\bigg],
\label{eq:bin-likelihood}
\end{equation} 
where $\rm N_{\rm bins}=14$ is the number of logarithmically uniform
bins in the range [28 TeV,10 PeV]. The numbers $n_{k,i}$ correspond to the {\em observed} events of topology $k$, shower or track, in the $i$-th bin. Thus, these numbers just depend on the assumption about the particular scenario considered for each analysis. On the other hand, the values $\mu_{k,i}(\theta)$ are the floating number of events of topology $k$ in the $i$-th bin. Accordingly, these numbers depend on the free parameters of the fit, which in general are given by $\theta=(N_a, \gamma,\, $f$_{\nu_\alpha(\bar{\nu}_\alpha)})$. 

Thus, the best fit point is obtained by minimizing eq. (\ref{eq:bin-likelihood}) with respect to the free paramters of the fit. Later, the confidence regions are computed by considering proper threshold values of the function $2\Delta \ln \lambda(\theta)= -2\ln \lambda(\theta) + 2\ln \lambda(\theta_{\text{best-fit}})$, which are tabulated in \cite{Agashe:2014kda} and depend on the number of free parameters. For instance, for the triangle plots we use $2\Delta \ln \lambda(\theta)=2.30,\,5.99$ and $6.63$ for the C.L. regions at $1\sigma,\,2\sigma$ and $3\sigma$ correspondingly.

Notice that during the analysis of the four scenarios considered in Section 3, which account for two different sources times two different exposures, we consider only one simulation of the {\em observed} data in each case. The number of events at each bin is given by the theoretical expected values obtained directly from the computation of flux times cross section times detector effects. 

Clearly, this procedure may be improved. For instance, we could consider multiple realizations of the data simulation by using poisson distributed numbers with mean values given by the theoretical predictions. Then, we could use the density maps containing the best fit point of each realization in order to find the global best fit and the confidence level regions. Also, we may consider the uncertainty on the numbers associated to the background predictions by including nuisance parameters into the definition of the likelihood. Indeed, we have checked both options and we have found that the general results of our analysis are not modified. The most clear effects are that the confidence level regions are not necessarily well shaped ellipses and cover slightly bigger areas, however they still keep the same basic features that we found with our simpler analysis.  
We believe that our simplified approach to show only the case of
the single realization is sufficient to illustrate the main point of this work.

\section{Astrophysical neutrinos }

\begin{figure}[htb]
\vglue -0.2cm
\centering
\begin{minipage}{1.0\textwidth}
\begin{minipage}{.5\textwidth}
\centering
\includegraphics[width=1.0\linewidth]{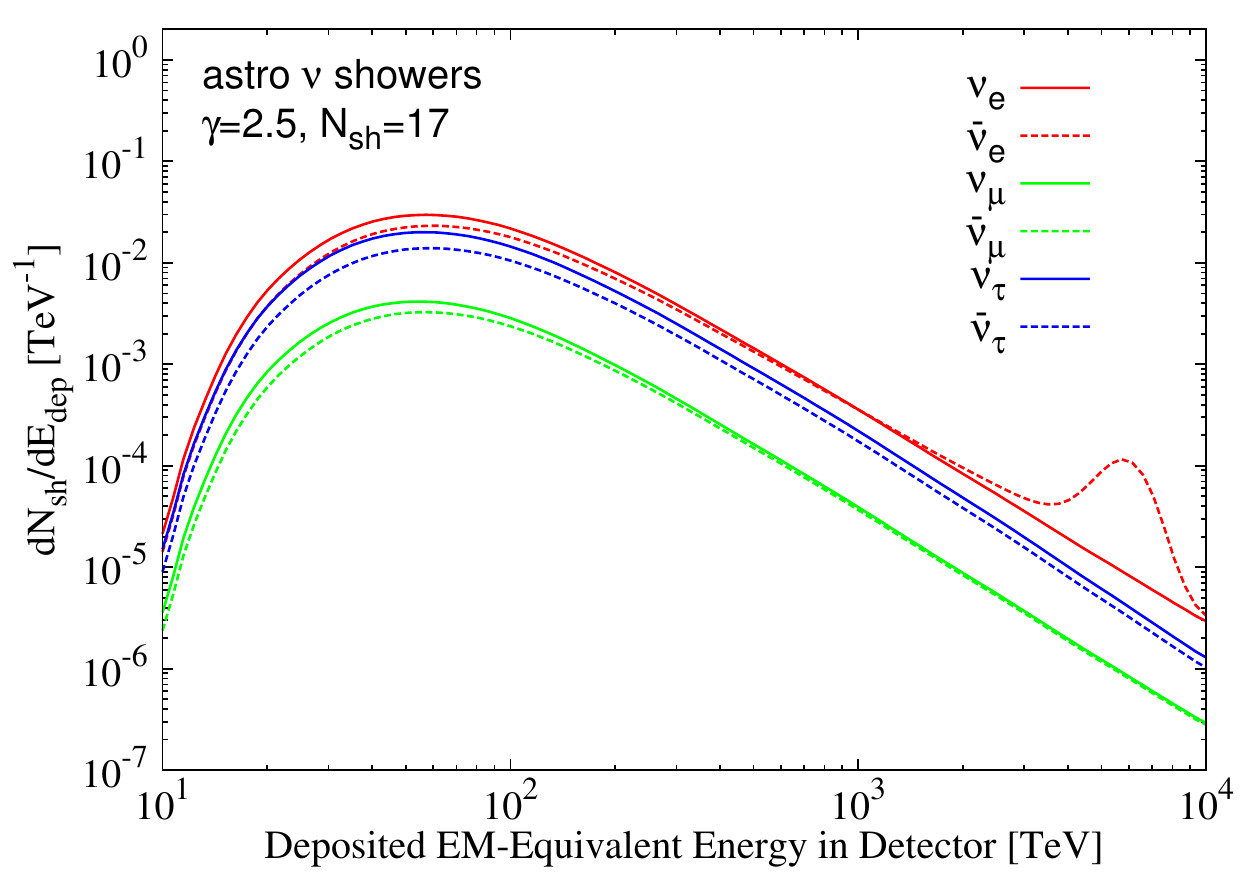}
\end{minipage}
\begin{minipage}{.5\textwidth}
\centering
\includegraphics[width=1.0\linewidth]{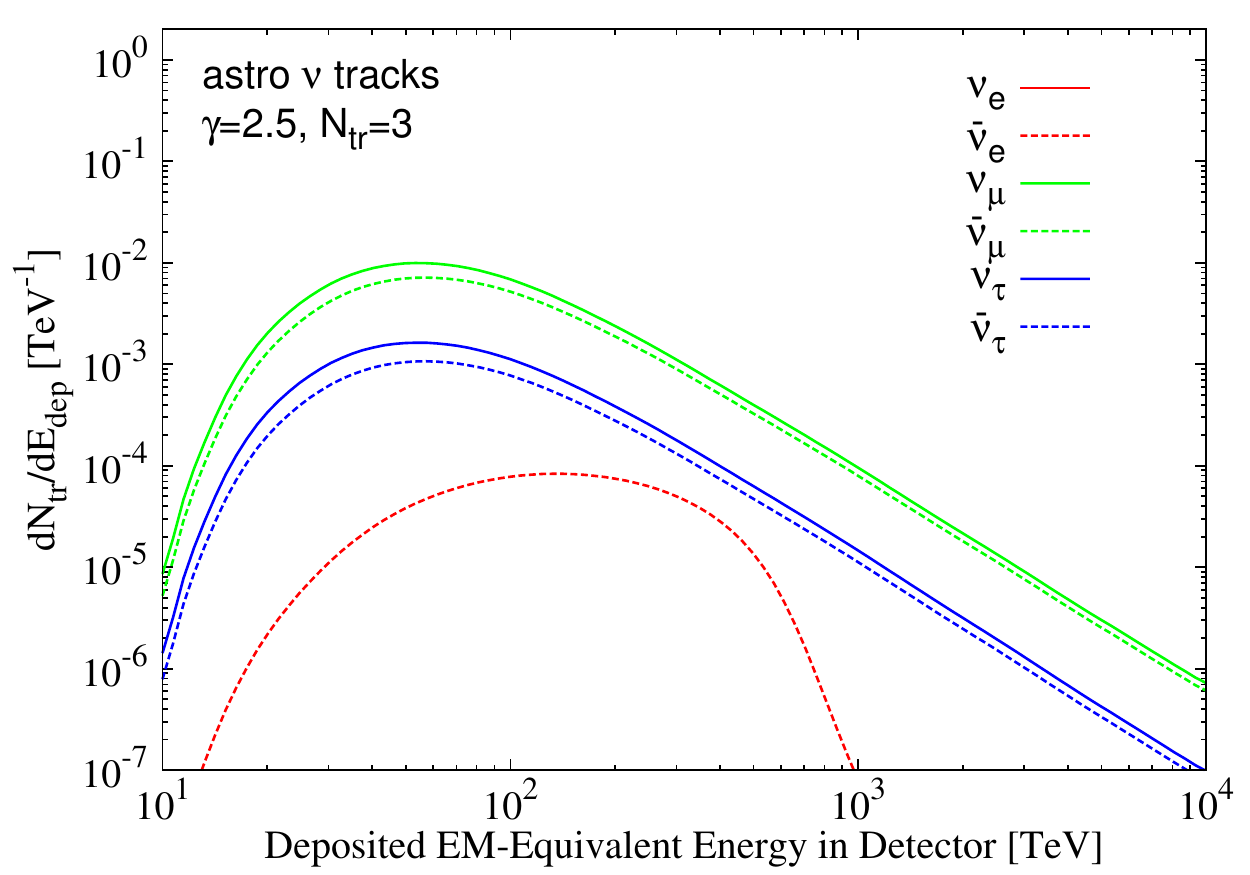}
\end{minipage}
\end{minipage}
\vglue 0.1cm
\caption{\label{fig6:Signal-contribution-to-ShTr}
Contribution of each flavor to shower (left panel) and track (right  panel) 
events induced by astrophysical neutrinos
as a function of the deposited energy in the IceCube detector, as simulated by us.}
\end{figure}

We assume that the flux of astrophysical neutrinos at Earth is given
by eq. (\ref{eq:fluxatearth}). In order to generate standard tables of shower and 
track events, we normalize the value
of $C_{r}$ such that 20 astrophysical neutrinos are obtained in the
interval $[10\,\text{TeV,}\,10\,\text{PeV]}$ and we consider that the
six independent neutrino fractions at Earth are equal to $1/6$. 
In Fig. \ref{fig6:Signal-contribution-to-ShTr}
we show the contributions of each flavor to the spectrum of showers
and tracks for $\gamma=2.5$ and $N_{a}=20$, where $N_{a}$ is the
total number of events with astrophysical origin. Notice that the electron-neutrino does not contribute to the tracks because there is no interaction channel that produces a muon or a tau. All the
rest can produce both showers and tracks. These results are
obtained from the convolution of each neutrino flux with the corresponding
neutrino-nucleon cross sections plus detector effects, for which we
have followed tightly the reference \cite{Palomares-Ruiz:2015mka} and references
therein. 

\section{High-energy neutrino background}

With respect to the background, we consider the contribution of
electron and muon atmospheric neutrinos plus the fake contribution
from atmospheric muons.  In the corresponding tables we have split
the contribution of neutrinos and anti-neutrinos. To compute the
electron and muon neutrino atmospheric flux we consider the
expressions given in \cite{Sinegovskaya:2014pia},
\begin{eqnarray}
\label{eq:conv-atm-flux}
\frac{d\Phi_{\alpha}(E_{\nu})}{dE} & = &
 C_{\alpha}\bigg(\frac{E_{\nu}}{1\,\text{GeV}}\bigg)^{-(g_{0}+g_{1}y+g_{2}y^{2})}\,\,\,\,\,\,\text{with\,\,\,}y=\frac{E_{\nu}}{1\,\text{GeV}}\ ,
\end{eqnarray}
where the values of $C_{\alpha}$, $g_{0}$, $g_{1}$ and $g_{2}$
can be found in this reference, for each flavor. 
We consider equal contributions from neutrinos and anti-neutrinos. 
In order to include the experimental veto for downgoing atmospheric muon-neutrinos 
we follow the algorithm showed in section III of \cite{Palomares-Ruiz:2015mka}, which is based on 
the following references \cite{Schonert:2008is,Gaisser:2014bja}. 

Similarly, for the muon flux we also follow~\cite{Palomares-Ruiz:2015mka}, 
so we consider the absolute number of tracks produced from the
atmospheric muon background as given by
\begin{eqnarray}
\frac{dN(E_{\nu})}{dE} & = & \begin{cases}
C_{\mu}(E_{\nu}/1\,\text{GeV})^{-\gamma_{\mu}} & \text{for}\, E>E_{\text{min}}\\
C_{\mu}(E_{\text{min}}/1\,\text{GeV})^{-\gamma_{\mu}-18.5}(E_{\nu}/1\,\text{GeV})^{18.5}\, & \text{for}\, E<E_{\text{min}}
\end{cases}
\label{eq:conv-muon-flux}
\end{eqnarray}
\noindent where 
\begin{eqnarray*}
\gamma_{\mu} & = & \frac{\log(21)}{(\log(60\,\text{TeV}/E_{\min}))}+1\,\,\,\text{and\,\,\,}E_{\text{min}}=28\,\text{TeV.}
\end{eqnarray*}

For each analysis showed in the body of the paper we fix the number of atmospheric neutrinos to $N_{\nu}=6.6$ and the number of atmospheric muons to $N_{\mu}=8.4$ in order to match the central values obtained by IceCube in the interval $[28\,\text{TeV,} 10\,\text{PeV]}$ after 988 days of data taking~\cite{Aartsen:2014gkd}. Notice that in the later reference these central values are reported with their respective uncertainties, however we do not consider this information in the paper. Nonetheless, we have checked that the effect of these uncertainties, included as nuisance parameters into the fit, do not affect our final conclusions. 

\begin{figure}[htb]
\vglue -0.2cm
\centering
\begin{minipage}{1.0\textwidth}
\begin{minipage}{.5\textwidth}
\centering
\includegraphics[width=1.0\linewidth]{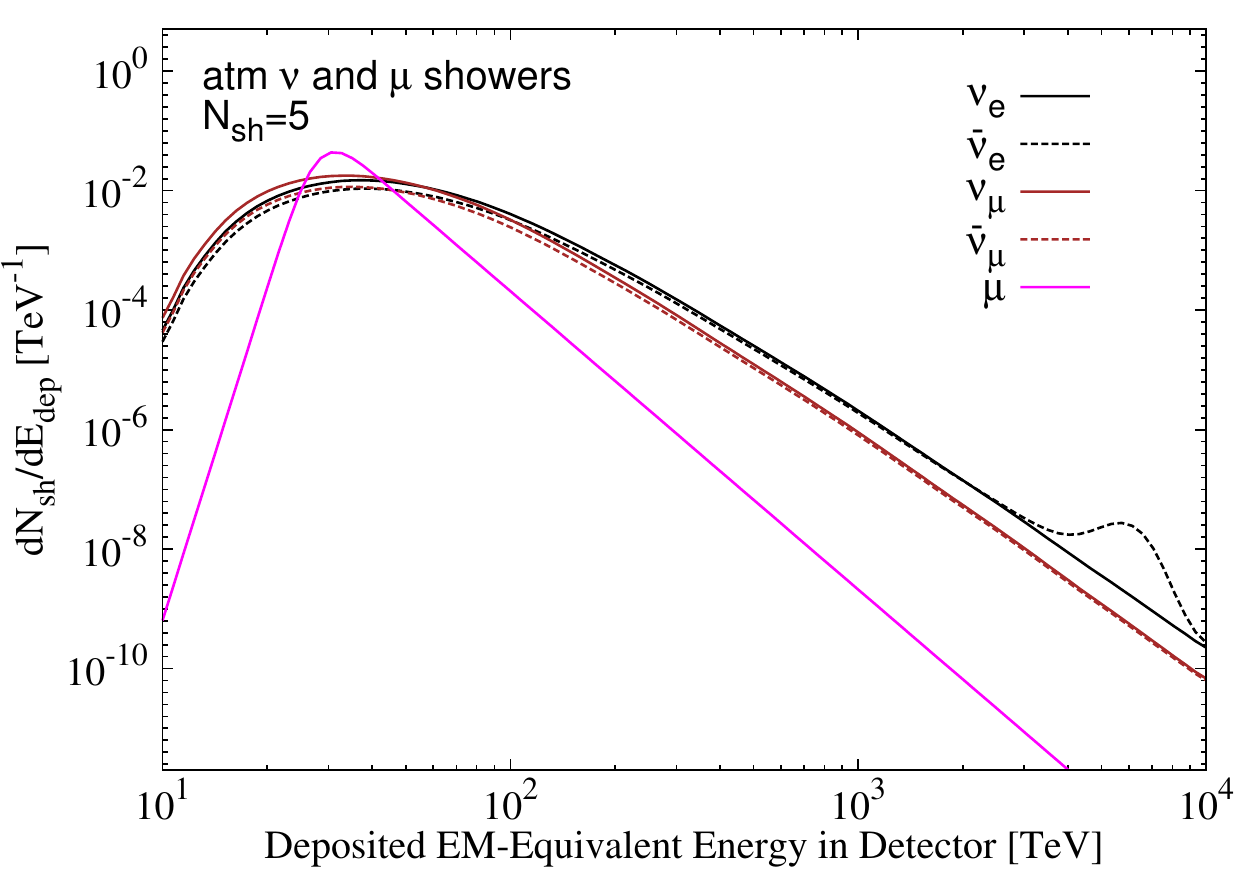}
\end{minipage}
\begin{minipage}{.5\textwidth}
\centering
\includegraphics[width=1.0\linewidth]{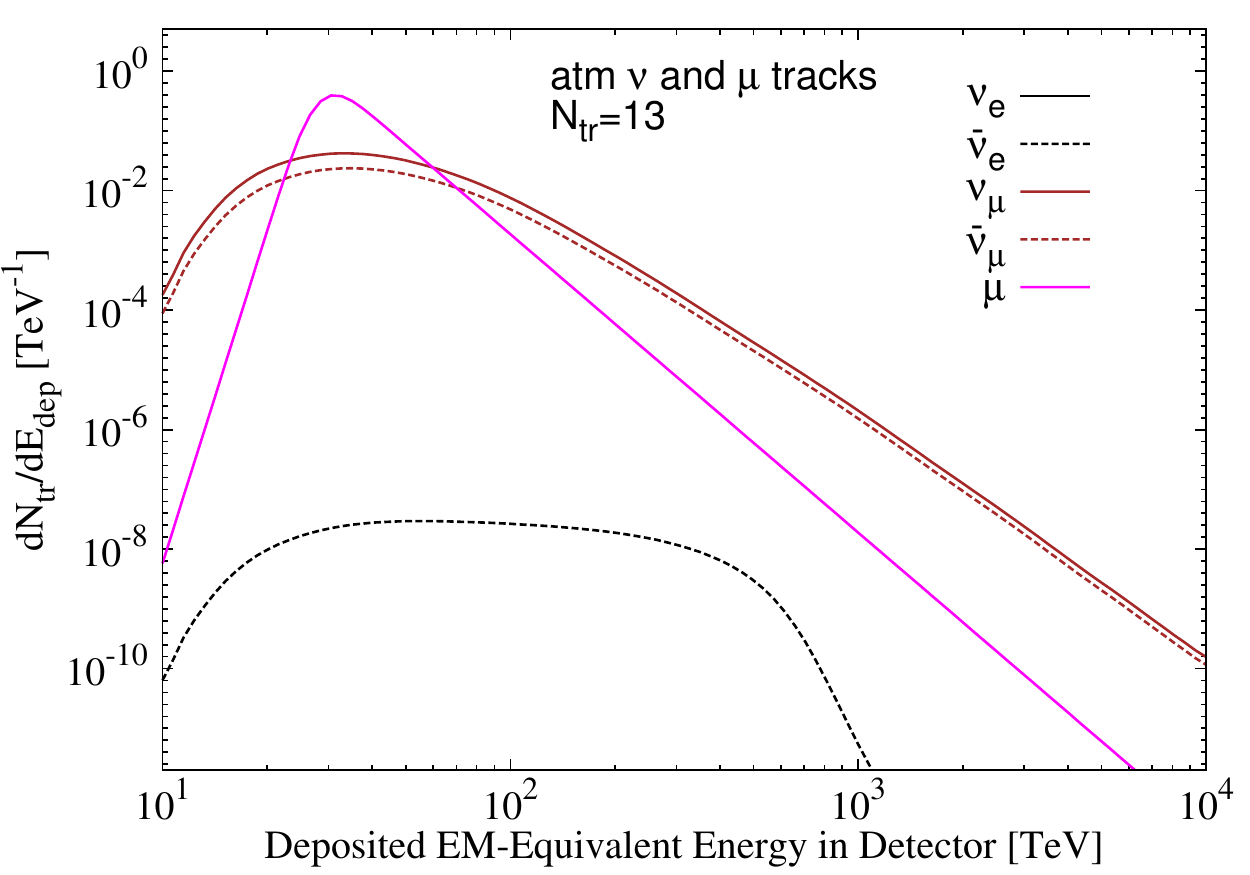}
\end{minipage}
\end{minipage}
\vglue 0.2cm
\caption{\label{fig7:Background-contribution-to-ShTr} 
Background contribution to showers (left panel)
and tracks (right panel). 
Notice that the y-axis has been extended to 
$10^{-14}\, {\rm GeV}^{-1}$ in order to cover interesting features of the
background.}

\end{figure}

We show the background contributions to shower and tracks in 
Fig.~\ref{fig7:Background-contribution-to-ShTr}. The neutrino curves are obtained by performing
the convolution of the fluxes given in eq. (\ref{eq:conv-atm-flux}) 
with the neutrino-nucleon cross sections in the same way as we do for astrophysical neutrinos.
Then, it is natural to observe some similarities between the neutrino spectra of Figs.~\ref{fig6:Signal-contribution-to-ShTr} and \ref{fig7:Background-contribution-to-ShTr}. The muon spectrum on the other hand is obtained directly from eq. (\ref{eq:conv-muon-flux}) and turns out to be quite distinct.

\newpage

\bibliographystyle{JHEP}
\bibliography{./icecube-nunubar-JCAP}

\end{document}